\documentclass[11pt]{article}
\pdfoutput=1
\usepackage{xspace,amsmath,amssymb,amsfonts}
\usepackage{times}
\usepackage{fullpage}
\usepackage{url}
\usepackage{graphicx,epsfig}

\newcommand{\ie}{\emph{i.e.}}
\newcommand{\eg}{\emph{e.g.}}

\newcommand{\DD}{{\cal D}}

\newcommand{\Sim}{\textsf{Sim}}
\newcommand{\Aux}{\textsf{Aux}}
\newcommand{\aux}{\textsf{aux}}
\newcommand{\wt}{\textsf{wt}}
\newcommand{\maxx}{\textsf{max}}
\newcommand{\minn}{\textsf{min}}
\newcommand{\supp}{\textsf{supp}}
\newcommand{\score}{\textsf{Score}}

\newtheorem{theorem}{Theorem}
\newtheorem{lemma}{Lemma}
\newtheorem{definition}{Definition}
\newcommand{\qed}{\hfill \ensuremath{\Box}}

\title{Robust De-anonymization of Large Datasets \\
({How to Break Anonymity of the Netflix Prize Dataset})}
\author{Arvind Narayanan and Vitaly Shmatikov \\[2ex]
        The University of Texas at Austin}

\begin{document}

\maketitle

\begin{abstract}

We present a new class of statistical de-anonymization attacks
against high-dimensional micro-data, such as individual preferences,
recommendations, transaction records and so on.  Our techniques are
robust to perturbation in the data and tolerate some mistakes in the
adversary's background knowledge.

We apply our de-anonymization methodology to the Netflix Prize dataset,
which contains anonymous movie ratings of 500,000 subscribers of Netflix,
the world's largest online movie rental service.  We demonstrate that
an adversary who knows only a little bit about an individual subscriber
can easily identify this subscriber's record in the dataset.  Using the
Internet Movie Database as the source of background knowledge, we
successfully identified the Netflix records of known users, uncovering
their apparent political preferences and other potentially sensitive
information.

\end{abstract}

\section{Introduction}

Datasets containing ``micro-data,'' that is, information about specific
individuals, are increasingly becoming public---both in response to ``open
government'' laws, and to support data mining research.  Some datasets
include legally protected information such as health histories; others
contain individual preferences, purchases, and transactions, which many
people may view as private or sensitive.

Privacy risks of publishing micro-data are well-known.  Even if
identifying information such as names, addresses, and Social Security
numbers has been removed, the adversary can use contextual and background
knowledge, as well as cross-correlation with publicly available databases,
to re-identify individual data records.  Famous re-identification
attacks include de-anonymization of a Massachusetts hospital discharge
database by joining it with with a public voter database~\cite{sweeney},
de-anonymization of individual DNA sequences~\cite{malin}, and privacy
breaches caused by (ostensibly anonymized) AOL search data~\cite{aol}.


Micro-data are characterized by high dimensionality and sparsity.
Informally, micro-data records contain many attributes, each of
which can be viewed as a dimension (an attribute can be thought
of as a column in a database schema).  Sparsity means that a pair
of random records are located far apart in the multi-dimensional
space defined by the attributes.  This sparsity is empirically
well-established~\cite{bryn,longtail,LAH06} and related to the ``fat
tail'' phenomenon: individual transaction and preference records tend
to include statistically rare attributes.

\vspace{1ex}
\noindent
\textbf{Our contributions.}
We present a very general class of statistical de-anonymization algorithms
which demonstrate the fundamental limits of privacy in public micro-data.
We then show how these methods can be used in practice to de-anonymize
the Netflix Prize dataset, a 500,000-record public dataset.

Our first contribution is a rigorous formal model for privacy breaches in
anonymized micro-data (section~\ref{model}).  We present two definitions,
one based on the probability of successful de-anonymization, the other
on the amount of information recovered about the target.  Unlike previous
work~\cite{sweeney}, we do not assume \emph{a priori} that the adversary's
knowledge is limited to a fixed set of ``quasi-identifier'' attributes.
Our model thus encompasses a much broader class of de-anonymization
attacks than simple cross-database correlation.

Our second contribution is a general de-anonymization algorithm
(section~\ref{meta-alg}).  Under very mild assumptions about the
distribution from which the records are drawn, the adversary with a small
amount of background knowledge about an individual can use it to identify,
with high probability, this individual's record in the anonymized dataset
and to learn all anonymously released information about him or her,
including sensitive attributes.  For \emph{sparse} datasets, such as
most real-world datasets of individual transactions, preferences, and
recommendations, very little background knowledge is needed (as few as
5-10 attributes in our case study).  Our de-anonymization algorithm is
\emph{robust} to imprecision of the adversary's background knowledge
and to sanitization or perturbation that may have been applied to the
data prior to release.  It works even if only a \emph{subset} of the
original dataset has been published.


Our third contribution is a practical analysis of the Netflix Prize
dataset, containing anonymized movie ratings of 500,000 Netflix
subscribers (section~\ref{sec:netflix}).  Netflix---the world's largest
online movie rental service---published this dataset to support the
Netflix Prize data mining contest.  We demonstrate that an adversary who
knows only a little bit about an individual subscriber can easily identify
his or her record if it is present in the dataset, or, at the very least,
identify a small set of records which include the subscriber's record.
The adversary's background knowledge need not be precise, \eg, the dates
may only be known to the adversary with a 14-day error, the ratings may
be known only approximately, and some of the ratings and dates may even
be completely wrong.  Because our algorithm is robust, if it uniquely
identifies a record in the published dataset, with high probability
this identification is not a false positive, even though the dataset
contains the records of only $\frac{1}{8}$ of all Netflix subscribers
(as of the end of 2005, which is the ``cutoff'' date of the dataset),

\section{Related work}

Unlike statistical databases~\cite{TYW84,AW89,AS00,CDM05,sulq},
micro-data datasets contain actual records of individuals even
after anonymization.  A popular approach to micro-data privacy is
$k$-anonymity~\cite{ksweeney,sweeney-supp,ciriani}.  The data publisher
must determine in advance which of the attributes are available to the
adversary (these are called ``quasi-identifiers''), and which are the
``sensitive attributes'' to be protected.  $k$-anonymization ensures
that each ``quasi-identifier'' tuple occurs in at least $k$ records
in the anonymized database.  It is well-known that $k$-anonymity does
not guarantee privacy, because the values of sensitive attributes
associated with a given quasi-identifier may not be sufficiently
diverse~\cite{ldiversity,icde07} or because the adversary has
access to background knowledge~\cite{ldiversity}.  Mere knowledge
of the $k$-anonymization algorithm may be sufficient to break
privacy~\cite{leizhang}.  Furthermore, $k$-anonymization completely
fails on high-dimensional datasets~\cite{curse}, such as the Netflix
Prize dataset and most real-world datasets of individual recommendations
and purchases.

In contrast to previous attacks on micro-data privacy~\cite{sweeney},
our de-anonymization algorithm does not assume that the attributes are
divided \emph{a priori} into quasi-identifiers and sensitive attributes.
Examples include anonymized transaction records (if the adversary
knows a few of the individual's purchases, can he learn \emph{all} of
her purchases?), recommendation and rating services (if the adversary
knows a few movies that the individual watched, can he learn \emph{all}
movies she watched?), Web browsing and search histories~\cite{aol},
and so on.  In such datasets, it is impossible to tell in advance
which attributes might be available to the adversary; the adversary's
background knowledge may even vary from individual to individual.
Unlike~\cite{sweeney,malin,lens}, our algorithm is \emph{robust}.
It works even if the published records have been perturbed, if only a
subset of the original dataset has been published, and if there are
mistakes in the adversary's background knowledge.

Our main case study is the Netflix Prize dataset of movie ratings.
We are aware of only one previous paper that considered privacy of movie
ratings.  In collaboration with the MovieLens recommendation service,
Frankowski \emph{et al.} correlated public mentions of movies in the
MovieLens discussion forum with the users' movie rating histories in the
\emph{internal} MovieLens dataset~\cite{lens}.  The algorithm uses the
entire public record as the background knowledge (29 ratings per user,
on average), and is not robust if this knowledge is imprecise (\eg,
if the user publicly mentioned movies which he did not rate).

By contrast, our analysis is based solely on public data.  Our
de-anonymization is \emph{not} based on cross-correlating Netflix internal
datasets (to which we do not have access) with public Netflix forums.
It requires much less background knowledge (2-8 ratings per user),
which need not be precise.  Furthermore, our analysis has privacy
implications for 500,000 Netflix subscribers whose records have been
published; by contrast, the largest public MovieLens datasets contains
only 6,000 records.

\section{Model}
\label{model}

\noindent
\textbf{Database.}
Define database $\DD$ to be an $N\times M$ matrix where each row is a record associated
with some individual, and the columns are attributes.  We are interested
in databases containing individual preferences and transactions, such as
shopping histories, movie or book preferences, Web browsing histories,
and so on.  Thus, the number of columns reflects the total number of
items in the space we are considering, ranging from a few thousands for
movies to millions for (say) the \url{amazon.com} catalog.

Each attribute (column) can be thought of as a dimension, and each
individual record as a point in the multidimensional attribute space.
To keep our analysis general, we will not fix the space $X$ from which
attributes are drawn.  They may be boolean (\eg, has this book been
rated), integer (\eg, the book's rating on a 1-10 scale), date, or a
tuple such as a (rating, date) pair.


The typical reason to publish anonymized databases is ``collaborative
filtering,'' \ie, predicting a customer's future choices from his past
behavior using the knowledge of what similar customers did.  Abstractly,
the goal is to predict the value of some attributes using a combination
of other attributes.  This is used in shopping recommender systems,
aggressive caching in Web browsers, and so on~\cite{collab}.


\vspace{1ex}
\noindent
\textbf{Sparsity and similarity.}
\label{sparsity}
Preference databases with thousands to millions of attributes are
necessarily \emph{sparse}, \ie, each individual record contains values
only for a small fraction of attributes.  We call these \emph{non-null}
attributes, and the set of non-null attributes the \emph{support} of
a record (denoted $\supp(r)$).  Null attributes are denoted $\perp$.
The \emph{support} of a column is defined analogously.  For example, the
shopping history of even the most profligate Amazon shopper contains only
a tiny fraction of all available items.  Even though points corresponding
to database records are very sparse in the attribute space, each record
may have dozens or hundreds of non-null attributes, making the database
truly high-dimensional.

The distribution of support sizes of attributes is typically heavy- or
\emph{long-tailed}, roughly following the power law~\cite{bryn,longtail}.
This means that although the supports of the columns corresponding to
``unpopular'' items are small, these items are so numerous that they make
up the bulk of the non-null entries in the database.  Thus, any attempt
to approximate the database by projecting it down to the most common
columns is bound to failure.  (The same effect causes $k$-anonymization
to completely fail on such databases~\cite{curse}.)

Unlike ``quasi-identifiers''~\cite{ksweeney,ciriani}, each attribute
typically has very little entropy for de-anonymization purposes.  In a
large database, for any except the rarest attributes, there are hundreds
of records with the same value of this attribute as any given record.
Therefore, it is \emph{not} a quasi-identifier.  At the same time,
knowledge that a particular individual has a certain attribute value
does reveal \emph{some} information about her record, since attribute
values and even the mere fact that the attribute is non-null vary from
record to record.


The similarity measure $\Sim$ is a function that maps a pair of attributes
(or more generally, a pair of records) to the interval $[0,1]$.  It captures the
intuitive notion of two values being ``similar.''  Typically, $\Sim$
on attributes will behave like a delta function.  For example, in our
analysis of the Netflix Prize dataset, $\Sim$ outputs 1 on a pair of
movies rated by different subscribers if and only if both the ratings
and the dates are within a certain threshold of each other (otherwise
it outputs 0).  

To define $\Sim$ over two records $r_1, r_2$, we ``generalize''
the cosine similarity measure:
$$
\Sim(r_1, r_2) = \frac{\sum \Sim(r_{1i}, r_{2i})}{|\supp(r_1) \cup \supp(r_2)|}
$$

\begin{definition}[Sparsity]
A database $D$ is $(\epsilon, \delta)$-sparse w.r.t. the
similarity measure $\Sim$ if 
$$\Pr_r[\Sim(r, r') > \epsilon\ \forall r' \neq r] \leq \delta$$
\end{definition}

As a real-world example, in appendix~\ref{appendix-sparsity} we show
that the Netflix Prize dataset is overwhelmingly sparse: for the vast
majority of records, there isn't a \emph{single} similar record in the
entire 500,000-record dataset.

\vspace{1ex}
\noindent
\textbf{Sanitization and sampling.}
Our de-anonymization algorithms are designed to work against databases
that have been anonymized and ``sanitized'' by their publishers.
The three main sanitization methods are perturbation, generalization, and
suppression~\cite{sweeney-supp,ciriani}.  Furthermore, the data publisher
may only release a (possibly non-uniform) sample of the database.
For example, he may attempt to $k$-anonymize the records, and then
release only one record out of each cluster of $k$ or more records.


If the database is released for collaborative filtering or similar
data mining purposes (as in the case of the Netflix Prize dataset), the
``error'' introduced by sanitization \emph{cannot} be large, otherwise
its utility will be lost.  We make this precise in our analysis.
Our definition of privacy breach (see below) allows the adversary to
identify not just his target's record, but \emph{any} record as long as
it is sufficiently similar (via $\Sim$) to the target and can thus be
used to determine its attributes with high probability.


From the viewpoint of our algorithm, there is no difference between
the perturbation of the published records and the imprecision of the
adversary's knowledge about his target.  In both cases, there is a small
discrepancy between an attribute value in the target's anonymous record
and her attribute value as known to the adversary.  Our de-anonymization
algorithm is designed to be robust to both.  Therefore, in the rest
of the paper we will treat perturbation applied to the data simply as
imprecision of the adversary's knowledge, and assume that the public,
anonymized sample ${\hat D}$ is an (arbitrary, unless otherwise specified)
subset of $D$.

\vspace{1ex}
\noindent
\textbf{Adversary model.}
The adversary's goal is to de-anonymize an anonymous record $r$ from the
public database.  To model this formally, we sample $r$ randomly from
$D$ and give a little bit of \emph{auxiliary information} or background
knowledge related to $r$ to the adversary.  It is restricted to a subset
of the (possibly imprecise, perturbed, or simply incorrect) values of
$r$'s attributes, modeled as an arbitrary probabilistic function $\Aux
\colon X^M \rightarrow X^M$.  The attributes given to the adversary
may be chosen uniformly from the support of $r$, or according to some
other probability distribution.  Given this auxiliary information and an
anonymized sample of the database, the adversary's goal is to reconstruct
attribute values of the entire record $r$.


In $k$-anonymity, there is a rigid division between demographic
quasi-identifiers and sensitive medical attributes (none of which are
known to the adversary)~\cite{ksweeney}.  By contrast, in our model if
the adversary happens to know the value of some attribute for his target,
it becomes part of his auxiliary information and thus an ``identifier''
(but only for \emph{this} individual's record).  If revealing the value
of some attribute violates some individual's privacy (this depends on
the specific record), then it is ``sensitive'' for this individual.

\vspace{1ex}
\noindent
\textbf{Privacy breach: formal definitions.}
\label{sec:breach}
What does it mean to de-anonymize a record $r$?  The naive answer is
to find the ``right'' anonymized record in the public sample ${\hat
D}$.  This is hard to capture formally, however, because it requires
assumptions about the data publishing process (\eg, what if ${\hat
D}$ contains two copies of every original record?).  Fundamentally,
the adversary's objective is \emph{privacy breach}: he wants to learn
as much as he can about $r$'s attributes that he doesn't already know.
We give two different (but related) formal definitions, because there
are two distinct scenarios for privacy breaches in large databases.

The first scenario is automated large-scale de-anonymization.  For every
record $r$ about which he has some information, the adversary must
produce a single ``prediction'' for all attributes of $r$.  An example
is the attack that inspired $k$-anonymity~\cite{sweeney}: taking the
demographic data from a voter database as auxiliary information, the
adversary joins it with the anonymized hospital discharge database
and uses the resulting combination to determine the values of medical
attributes for each person who appears in both databases.

\begin{definition}
\label{def:anp}
A database $D$ can be $(\theta, \omega)$-deanonymized w.r.t. auxiliary
information $\Aux$ if there exists an algorithm $A$ which, on inputs $D$
and $\Aux(r)$ where $r \leftarrow D$ outputs $r'$ such that 
$$\Pr[\Sim(r, r') \geq \theta] \geq \omega$$
\end{definition}

Definition~\ref{def:anp} can be interpreted as an \emph{amplification
of background knowledge}: the adversary starts with $\aux=\Aux(r)$
which is close to $r$ on a small subset of attributes, and uses this
to compute $r'$ which is close to $r$ on the entire set of attributes.
This captures the \textbf{adversary's ability to gain information about
his target record}.  He does not need to precisely identify the ``right''
anonymized record (as we argue above, this notion may be meaningless).
As long the adversary finds \emph{some} record which is guaranteed to
be very similar to the target record, \ie, contains the same or similar
attribute values, privacy breach has occurred.

If operating on a sample of the entire database, the de-anonymization
algorithm must also detect whether its target record is part of
the sample, or has not been released at all.  In the following, the
probability is taken over the randomness of the sampling of $r$ from
${\hat D}$, $\Aux$ and $A$ itself.

\begin{definition}[De-anonymization]
\label{def:anps}
An arbitrary subset ${\hat D}$ of a database $D$ can be $(\theta,
\omega)$-deanonymized w.r.t. auxiliary information $\Aux$ if there
exists an algorithm $A$ which, on inputs ${\hat D}$ and $\Aux(r)$ where
$r \leftarrow D$
\begin{itemize}
\item
If $ r \in {\hat D}$, outputs 
$r'$ s.t. $\Pr[\Sim(r, r') \geq \theta] \geq \omega$
\item
if $r \notin {\hat D}$, outputs $\perp$ with probability at least $\omega$
\end{itemize}
\end{definition}

The same error threshold ($1-\omega$) is used for both ``failures''--false
positives and false negatives--in the above definition because the
parameters of the algorithm can be adjusted so that both rates are equal;
this is the so called ``equal error rate.''


In the second privacy breach scenario, the adversary produces a set or
``lineup'' of candidate records that include his target record $r$,
either because there is not enough auxiliary information to identify $r$
in the lineup or because he expects to perform additional, possibly
manual analysis to complete de-anonymization.   This is similar to
communication anonymity in mix networks~\cite{SD02}.

The \emph{number} of candidate records is not a good metric, because some
of the records may be much likelier candidates than others.  Instead,
we consider the probability distribution over the candidate records, and
use the conditional \emph{entropy} of $r$ given $\aux$ as the metric.
In the absence of an ``oracle'' to identify the target record $r$
in the lineup, the entropy of the distribution itself can be used as
a metric~\cite{SD02,DSCP02}.  If the adversary has such an ``oracle''
(this is a technical device used to measure the adversary's success; in
the real world, the adversary may not have an oracle telling him whether
de-anonymization succeeded), then privacy breach can be quantified by
answering a very specific question: \emph{how many bits of additional
information does the adversary need in order to output a record which
is similar to his target record?}

Thus, suppose that after executing the de-anonymization algorithm,
the adversary outputs records $r'_1, \ldots r'_k$ and the coresponding
probabilities $p_1, \ldots p_k$.  The latter can be viewed as an
\emph{entropy encoding} of the candidate records.  According to Shannon's
source coding theorem, the optimal code length for record $r'_i$
is $-\log p_i$.  We denote by $H_S(\Pi,x)$ this Shannon entropy of a
record $x$ w.r.t. a probability distribution $\Pi$.  In the following,
the expectation is taken over the coin tosses of $A$, the sampling of $r$
and $\Aux$.

\begin{definition}
[Entropic de-anonymization]
\label{def:ane}
A database $D$ can be $(\theta, H)$-deanonymized w.r.t. auxiliary
information $\Aux$ if there exists an algorithm $A$ which, on inputs $D$
and $\Aux(r)$ where $r \leftarrow D$ outputs a set of candidate records
$D'$ and probability distribution $\Pi$ such that
$$
E[\textsf{min}_{r' \in D', \Sim(r, r') \geq \theta}H_S(\Pi, r')] \leq H
$$
\end{definition}

This definition measures the minimum Shannon entropy of the candidate
set of records which are similar to the target record.  As we will show,
in sparse databases this set is likely to contain a single record,
thus taking the minimum is but a syntactic requirement.

When the minimum is taken over an empty set, we define it to be $H_0 =
\log_2 N$, the \emph{a priori} entropy of the target record.  Intuitively,
this models the adversary outputting a random record from the entire
database when he cannot compute a lineup of plausible candidates.
Formally, the adversary's algorithm $A$ can be converted into an
algorithm $A'$, which outputs the mean of two probability distributions:
one is the output of $A$, the other is the uniform distribution over $D$.
Observe that for $A'$, the minimum is always taken over a non-empty set,
and the expectation for $A'$ differs from that for $A$ by at most 1 bit.

\section{De-anonymization algorithm}
\label{meta-alg}

We start by describing a meta-algorithm or an algorithm template.
The inputs are a sample ${\hat D}$ of database $D$ and auxiliary
information $\aux = Aux(r), r \leftarrow D$.  The output is either a
record $r' \in {\hat D}$, or a set of candidate records and a probability
distribution over those records (following Definitions~\ref{def:anps}
and~\ref{def:ane}, respectively).  The three main components of the
algorithm are the scoring function, matching criterion, and record
selection.

The \textbf{scoring function} $\score$ assigns a numerical score to
each record in ${\hat D}$ based on how well it matches the adversary's
auxiliary information $\Aux$.  The \textbf{matching criterion} is the
algorithm that the adversary applies to the set of scores assigned
by the scoring function to determine if there is a match.  Finally,
\textbf{record selection} selects one ``best-guess'' record or
a probability distribution, if needed.

The template for the de-anonymization algorithm is as follows: 

\begin{enumerate}

\item
The adversary computes $\score(\aux, r')$ for each $r' \in {\hat D}$.

\item
The adversary applies the matching criterion to the resulting set of
scores and computes the matching set; if the matching set is empty,
the adversary outputs $\perp$ and exits. (This step can be skipped if
${\hat D}=D$, \ie, if the entire database has been released, or if the
adversary knows for certain that the target record has been released,
\ie, $r \in {\hat D}$).

\item
If the adversary's ``best guess'' is required (de-anonymization according
to Definitions~\ref{def:anp} and~\ref{def:anps}), the adversary outputs
$r' \in {\hat D}$ with the highest score.  If a probability distribution
over candidate records is required (de-anonymization according to
Definition~\ref{def:ane}), the adversary computes some non-decreasing
probability distribution based on the score and outputs this distribution.

\end{enumerate}

\noindent
\textbf{Algorithm 1A.}
The following simple instantiation of the above algorithm template
is sufficiently tractable to be formally analyzed in the rest of this
section.

\begin{itemize}

\item
$\score(\aux, r') = \minn_{i \in \supp(\aux)} \Sim(\aux_i, r'_i)$,
\ie, the score of a candidate record is determined by the least similar
attribute between it and the adversary's knowledge.

\item
The adversary computes the matching set 
$D' = \{r' \in {\hat D} : \score(\aux, r') > \alpha\}$ 
for some fixed constant $\alpha$.
The matching criterion is that $D'$ be nonempty.

\item
Probability distribution is uniform on $D'$.

\end{itemize}

\vspace{1ex}
\noindent
\textbf{Algorithm 1B.}
This algorithm incorporates several heuristics which have proved useful
in practical analysis (see section~\ref{sec:netflix}).  First, the
scoring function gives higher weight to statistically rare attributes.
The intuition is as follows: if the auxiliary information tells the
adversary that his target has a certain rare attribute, this contains
much more information for de-anonymization purposes than the knowledge
of a common attribute (\eg, it is more useful to know that the target
has purchased ``The Dedalus Book of French Horror'' than the fact that
she purchased a Harry Potter book).

Second, to improve robustness of the algorithm, the matching criterion
requires that the top score be significantly above the second-best score.
This measures how much the identified record ``stands out'' from other
candidate records.

\begin{itemize}

\item
$\score(\aux,r') =  \sum_{i \in \supp(\aux)} \wt(i) \Sim(\aux_i, r'_i)$
where $\wt(i) = \frac 1 {\log |\supp(i)|}$.

\item
The adversary computes
$\maxx = \maxx(S), \maxx_2 = \maxx_2(S)$ and
$\sigma = \sigma(S)$ where $S = \{\score(\aux, r'): r' \in {\hat D}\}$,
\ie, the highest and second-highest scores and the standard deviation
of the scores.
If $\frac{\maxx-\maxx_2}\sigma < \phi$, where $\phi$ is a fixed parameter 
called the \emph{eccentricity},
then there is no match; otherwise, the matching set consists of the
record with the highest score.

\item
Probability distribution is 
$\Pi(r') = c \cdot e^{\frac{\score(\aux, r')}\sigma}$ for each $r'$,
where $c$ is a constant that makes the distribution sum up to 1.
This weighs each matching record in inverse proportion to the likelihood
that the match in question is a statistical fluke.

\end{itemize}

\subsection{Analysis: general case}
\label{analysis-general}

We now quantify the amount of auxiliary information needed to de-anonymize
an arbitrary multi-dimensional dataset using our Algorithm 1A.  The
smaller the required auxiliary information (\ie, the fewer attribute
values the adversary needs to know about his target), the easier the
attack.

We start with the worst-case analysis and calculate how much
auxiliary information is needed in the most general case, without
any assumptions about the distribution from which the data are drawn.
In section~\ref{analysis-sparse}, we will show that much less auxiliary
information is needed to de-anonymize records drawn from \emph{sparse}
distributions (all known real-world examples of transactions and
recommendation datasets are sparse).

Let $\aux$ be the auxiliary information about some record $r$ from the
dataset.  This knowledge consists of $m$ (non-null) attribute values,
which are close to the corresponding values of attributes in $r$,
that is, $|\aux|=m$ and $\Sim(\aux_i, r_i) \geq 1-\epsilon\ \forall
i \in \supp(\aux)$, where $\aux_i$ (respectively, $r_i$) is the $i$th
attribute of $\aux$ (respectively, $r$).

\begin{theorem}\label{thmstrong}
Let $0 < \epsilon, \delta < 1$ and let $D$ be the database.  Let $\Aux$ be such
that
$\aux = \Aux(r)$ consists of at least $m \geq \frac{\log N-\log \epsilon}{-\log (1-\delta)}$ 
randomly selected attribute values of the target record $r$, with
$\Sim(\aux_i, r_i) \geq 1-\epsilon\ \forall i \in \supp(\aux)$.
Then $D$ can be
$(1-\epsilon-\delta,1-\epsilon)$-deanonymized w.r.t. $\Aux$.
\end{theorem}

{\bf Proof.} The adversary uses Algorithm 1A  with $\alpha=1-\epsilon$
to compute the set of all records in ${\hat D}$ that match $\aux$.
The adversary then outputs a record $r'$ at random from the matching set.
It is sufficient to prove that this randomly chosen $r'$ must be very
similar to the target record $r$. (This satisfies our definition of a
privacy breach because it gives to the adversary almost everything
he may want to learn about $r$.)

Record $r'$ is a \emph{false match} if $\Sim(r, r') \leq 1 - \epsilon -
\delta$ (\ie, the likelihood that it is similar to the target record $r$
is below the threshold).  We first show that, with high probability,
the matching set does not contain any false matches.

\begin{lemma}
\label{falsematchlemma}
If $r'$ is a false match, then $\Pr_{i \in \supp(r)}[\Sim(r_i, r_i')
\geq 1-\epsilon] < 1-\delta$
\end{lemma}

Lemma~\ref{falsematchlemma} holds, because the contrary implies
$\Sim(r,r') \geq (1-\epsilon)(1-\delta) \geq (1-\epsilon-\delta)$,
contradicting the assumption that $r'$ is a false match.  Therefore, the
probability that the false match $r'$ belongs to the matching set is at
most $(1-\delta)^m$.  By a union bound, the probability that there is even
a single false match $r'$ in the matching set is at most $N(1-\delta)^m$.
If $m = \frac{\log \frac{N}{\epsilon}}{\log \frac{1}{1-\delta}}$, then
the probability that the matching set returned by Algorithm 1A
contains any false matches is no more than $\epsilon$.

Therefore, with probability $1-\epsilon$, there are no false matches.
Thus for every record $r'$ in the matching set, $\Sim(r, r') \geq
1-\epsilon-\delta$, \ie, any $r'$ must be similar to the true record $r$.
To complete the proof, observe that the matching set contains at least
one record, $r$ itself.

When $\delta$ is small, $m=\frac{\log N-\log \epsilon}\delta$.  Note the
logarithmic dependence on $\epsilon$ and the linear dependence on
$\delta$: the chance that the de-anonymization algorithm completely
fails is very small even if attribute-wise accuracy is not very high.
Also note that the size of the matching set need not be small.  Even if
the de-anonymization algorithm returns a large number of records, with
high probability they are \emph{all} similar to the target record $r$,
and thus any one of them can be used to learn the unknown sensitive
attributes of $r$.

\subsection{Analysis: sparse datasets}
\label{analysis-sparse}

As shown in section~\ref{sparsity}, most real-world datasets containing
individual transactions, preferences, and so on are \emph{sparse}.
Sparsity increases the probability that de-anonymization succeeds,
decreases the amount of auxiliary information needed, and makes the
algorithm more robust to both perturbation in the data and mistakes in
the auxiliary information.  

Our assumptions about data sparsity are very mild.  We only assume
$(1-\epsilon-\delta,\ldots)$ sparsity, \ie, we assume that the average
record does not have \emph{extremely} similar peers in the dataset
(real-world records tend not to have even \emph{approximately} similar
peers---see appendix~\ref{appendix-sparsity}).

\begin{theorem}\label{thmmid}
Let $\epsilon$, $\delta$, and $\aux$ be as in Theorem~\ref{thmstrong}.
If the database $D$ is $(1-\epsilon-\delta,\epsilon)$-sparse, then $D$
can be $(1, 1-\epsilon)$-deanonymized.\qed
\end{theorem}

The proof is essentially the same as the proof of Theorem~\ref{thmstrong},
but in this case \emph{any} $r' \neq r$ from the matching set must be a
false match.  Because with probability $1-\epsilon$, Algorithm 1A
outputs no false matches, the matching set consists of exactly one record:
the true target record $r$.

Finally, de-anonymization in the sense of Definition~\ref{def:ane}
requires even less auxiliary information.  Recall that in this kind
of privacy breach, the adversary outputs a ``lineup'' of $k$ suspect
records, one of which is the true record.  By analogy with $k$-anonymity,
we will call this $k$-deanonymization.  Formally, this is equivalent
to $(1,\frac{1}{k})$-deanonymization in our framework.

\begin{theorem}
\label{thmweak}
Let $D$ be $(1-\epsilon-\delta, \epsilon)$-sparse and
$\aux$ be as in Theorem~\ref{thmstrong} with $m = \frac{\log
\frac{N}{k-1}}{\log\frac{1}{1-\delta}}$.  
Then \begin{itemize}
\item$D$ can be $(1, \frac 1 k)$-deanonymized.
\item$D$ can be $(1, \log k)$-deanonymized (entropically).
\end{itemize}
\end{theorem}

By the same argument as in the proof of Theorem~\ref{thmstrong},
if the number of attributes known to the adversary $m = \frac{\log
\frac{N}{k-1}}{\log\frac{1}{1-\delta}}$, then the expected number
of false matches in the set of records output by the de-anonymization
algorithm is at most $k-1$.  Let $X$ be the random variable representing
the number of false matches.  If the adversary outputs a random record
from the matching set, the probability of hitting a non-false match is at
least $\frac 1 X$.  Since $\frac 1 x$ is a convex function, we can apply
Jensen's inequality~\cite{jensen} to obtain $E[\frac 1 X] \geq \frac 1
{E(X)} \geq \frac 1 k$, resulting in $(1, \frac 1 k)$-deanonymization.

Similarly, if the adversary outputs the uniform distribution over
the matching set, then the entropy of de-anonymization is $\log X$.
But since $\log x$ is a concave function, by Jensen's inequality we
have $E[\log X] \leq \log E(X) \leq \log k$, resulting in $(1, \log k)$
entropic deanonymization. \qed

Note that neither assertion of the theorem follows directly from the
other.

\subsection{Analysis: de-anonymization from a sample}

We now consider the scenario in which the released database ${\hat D}
\subsetneq D$ is a sample of the original database $D$, \ie, only some
of the anonymized records are available to the adversary.  This is the
case, for example, for the Netflix Prize dataset (the subject of our
case study in section~\ref{sec:netflix}), where the publicly available
anonymized sample contains roughly $\frac{1}{8}$ of the original records.

In this scenario, even though the original database $D$ contains
the adversary's target record $r$, this record may not appear in
${\hat D}$ even in anonymized form.  The adversary can still apply
the de-anonymization algorithm, but there is a possibility that the
matching set is empty, in which case the adversary outputs $\perp$
(indicating that de-anonymization fails).  If the matching set is not
empty, he proceeds as before: picks a random record $r'$ and learn
the attributes of $r$ on the basis of $r'$.  We now demonstrate the
equivalent of Theorem~\ref{thmstrong}: de-anonymization succeeds as long
as $r$ is in the public sample; otherwise, the adversary can detect,
with high probability, that $r$ is not in the public sample.

\begin{theorem}
\label{thmsample}
Let $\epsilon$, $\delta$, $D$, and $\aux$ be as in Theorem
\ref{thmstrong}, and ${\hat D} \subset D$.  Then ${\hat D}$ can be
$(1-\epsilon-\delta, 1-\epsilon)$-deanonymized w.r.t. $\aux$. \qed
\end{theorem}

The bound on the probability of a false match given in the proof of
Theorem~\ref{thmstrong} still holds, and the adversary is guaranteed
at least one match as long as his target record $r$ is in ${\hat D}$.
Therefore, if $r \notin {\hat D}$, the adversary outputs $\perp$ with
probability at least $1-\epsilon$.  If $r \in {\hat D}$, then again the
adversary succeeds with probability at least $1-\epsilon$.

On the other hand, theorems~\ref{thmmid} and~\ref{thmweak} do not
directly translate to this scenario.  For each record in the public sample
${\hat D}$, there could be an arbitrary number of similar records in $D
\setminus {\hat D}$ (\ie, the part of the database that is not available
to the adversary).  


Fortunately, if $D$ happens to be sparse in the sense of
section~\ref{sparsity} (recall that all real-world databases are sparse
in this sense), then theorems~\ref{thmmid} and~\ref{thmweak} still hold,
that is, de-anonymization succeeds with a very small amount of auxiliary
information.  The following theorem ensures that if the random sample
${\hat D}$ available to the adversary is sparse, then the entire database
$D$ must also be sparse.  Therefore, the adversary can simply apply the
de-anonymization algorithm to the sample, and be assured that if the
target record $r$ has been de-anonymized, then with high probability
this is not a false positive.

\begin{theorem}
\label{samplesparsity}
If database $D$ is not $(\epsilon, \delta)$-sparse, then a random
$\frac{1}{\lambda}$-subset $\hat D$ is not $(\epsilon, \frac
{\delta\gamma}\lambda)$-sparse with probability at least $1-\gamma$. \qed
\end{theorem}

For each $r \in \hat D$, the ``nearest neighbor'' $r'$ of $r$ in $D$
has a probability $\frac 1 \lambda$ of being included in $\hat D$.
Therefore, the expected probability that the similarity with the nearest
neighbor is at least $1-\epsilon$ is at least $\frac \delta \lambda$.
(Here the expectation is over the set of all possible samples and the
probabiility is over the choice of the record in $\hat D$.)  Applying
Markov's inequality, the probability, taken over the choice $\hat D$, that
${\hat D}$ is sparse, \ie, that the similarity with the nearest neighbor
is $\frac{\delta\gamma}{\lambda}$, is no more than $\gamma$.  \qed

The above bound is quite pessimistic.  Intuitively, for any ``reasonable''
dataset, the sparsity of a random sample will be about the same as that
of the original dataset.

Theorem~\ref{samplesparsity} can be interpreted as follows.  Consider
the adversary who has access to a sparse sample $\hat D$, but not the
entire database $D$.  Theorem~\ref{samplesparsity} says that either
a very-low-probability event has occurred, or $D$ itself is sparse.
Note that it is meaningless to try to bound the probability that $D$
is sparse because we do not have a probability distribution on how $D$
itself is created.

Intuitively, this implies that unless the sample is specially tailored,
sparsity of the sample implies sparsity of the entire database.
The alternative is that the similarity with the nearest neighbor of a
random record in the sample is very different from the corresponding
distribution in the full database.

In practice, most, if not all anonymized real-world datasets and samples
are published to support research on data mining and collaborative
filtering (this is certainly the case for the Netflix Prize dataset,
which is the subject of our case study in section~\ref{sec:netflix}).
Tailoring the published sample in such a way that the nearest-neighbor
similarity in the sample is radically different from that in the
original dataset would completely destroy utility of the sample for
learning new collaborative filters, which is often based on the set of
nearest neighbors.  Therefore, in real-world anonymous data publishing
scenarios---including the Netflix Prize dataset---sparsity of the sample
should imply sparsity of the original dataset.

\section{Case study: Netflix Prize dataset}
\label{sec:netflix}

On October 2, 2006, Netflix, the world's largest online DVD rental
service, announced the \$1-million Netflix Prize for improving their movie
recommendation service~\cite{nyt}.  To aid contestants, Netflix publicly
released a dataset containing $100,480,507$ movie ratings, created by
$480,189$ Netflix subscribers between December 1999 and December 2005.
At the end of 2005, Netflix had approximately 4 million subscribers, so
almost $\frac{1}{8}$ of them had their records published.  The ratings
data appear to not have been perturbed to any significant extent (see
appendix~\ref{liars}).

While movie ratings are not as sensitive as, say, medical records,
release of massive amounts of data about individual Netflix subscribers
raises interesting privacy issues.  Among the Frequently Asked Questions
on the Netflix Prize webpage~\cite{faq}, there is the following question:
``Is there any customer information in the dataset that should be kept
private?''  Netflix answers this question as follows:

\begin{quote}
``No, all customer identifying information has been removed; all that remains are ratings and dates. This follows our privacy policy, which you can review here. Even if, for example, you knew all your own ratings and their dates you probably couldn't identify them reliably in the data because only a small sample was included (less than one-tenth of our complete dataset) and that data was subject to perturbation. Of course, since you know all your own ratings that really isn't a privacy problem is it?''
\end{quote}

Of course, removing the identifying information from the records is not
sufficient for anonymity.  An adversary may have auxiliary information
about some subscriber's movie preferences: the titles of a few of the
movies that this subscriber watched, whether she liked them or not, maybe
even approximate dates when she watched them.  Anonymity of the Netflix
dataset thus depends on the answer to the following question: \textbf{How
much does the adversary need to know about a Netflix subscriber in order
to identify her record in the dataset, and thus learn her complete movie
viewing history?}

In the rest of this section, we investigate this question.  Formally,
we will study the numerical relationship between the size of $\Aux$
and $(1,\omega)$- and $(1,H)$-deanonymization.

\vspace{1ex}
\noindent
\textbf{Does privacy of Netflix ratings matter?}
The privacy question is \emph{not} ``Does the average Netflix
subscriber care about the privacy of his movie viewing history?,''
but ``Are there \emph{any} Netflix subscribers whose privacy can be
compromised by analyzing the Netflix Prize dataset?''  The answer to
the latter question is, undoubtedly, yes.  As shown by our experiments
with cross-correlating non-anonymous records from the Internet Movie
Database with anonymized Netflix records (see below), it is possible to
learn sensitive \emph{non-public} information about a person's political
or even sexual preferences.  We assert that even if the vast majority
of Netflix subscribers did not care about the privacy of their movie
ratings (which is not obvious by any means), our analysis would
still indicate serious privacy issues with the Netflix Prize dataset.

Moreover, the linkage between an individual and her movie viewing history
has implications for her \emph{future} privacy.  In network security,
``forward secrecy'' is important: even if the attacker manages to
compromise a session key, this should not help him much in compromising
the keys of future sessions.  Similarly, one may state the ``forward
privacy'' property: if someone's privacy is breached (\eg, her anonymous
online records have been linked to her real identity), future privacy
breaches should not become easier.  Now consider a Netflix subscriber
Alice whose entire movie viewing history has been revealed.  Even if in
the future Alice creates a brand-new virtual identity (call her Ecila),
Ecila will \emph{never} be able to disclose any non-trivial information
about the movies that she had rated within Netflix because any such
information can be traced back to her real identity via the Netflix
Prize dataset.  In general, once any piece of data has been linked to
a person's \emph{real} identity, any association between this data and
a \emph{virtual} identity breaks anonymity of the latter.

It also appears that Netflix might be in violation of its own stated
privacy policy.  According to this policy, ``Personal information
means information that can be used to identify and contact you,
specifically your name, postal delivery address, e-mail address,
payment method (e.g., credit card or debit card) and telephone number,
as well as other information when such information is combined with your
personal information.  [...]  We also provide analyses of our users in the
aggregate to prospective partners, advertisers and other third parties. We
may also disclose and otherwise use, on an anonymous basis, movie ratings,
commentary, reviews and other non-personal information about customers.''
The simple-minded division of information into personal and non-personal
is a false dichotomy.

\vspace{1ex}
\noindent
\textbf{Breaking anonymity of the Netflix Prize dataset.}
We apply our Algorithm 1B from section~\ref{meta-alg} to the Netflix
Prize dataset.  The similarity measure $\Sim$ on attributes is a threshold
function: $\Sim$ returns 1 if and only if the two attribute values are
within a certain threshold of each other.  For the rating attributes,
which in the case of the Netflix Prize dataset are on the 1-5 scale, we
consider the thresholds of 0 (corresponding to exact match) and 1, and
for the date attributes, 3 and 14 days. We also consider the threshold
of $\infty$ for the date, which models the adversary having not being
given any dates as part of the auxiliary information.

In addition, we allow some of the attribute values in the attacker's auxiliary
information to be completely wrong. Thus, we say that $\aux$ of a
record $r$ consists of $m$ movies out of $m'$ if $|\aux| = m'$ and 
$\sum \Sim(\aux_i, r_i) \geq m$.
We instantiate the scoring function as follows:
$$
\score(\aux,r') = {\sum_{i \in \supp(\aux)} \wt(i)(e^{\frac{\rho_i-\rho'_i}{\rho_0}}+e^{\frac{d_i-d'_i}{d_0}})}
$$
where $\wt(i) = \frac 1 {\log |\supp(i)|}$ ($|\supp(i)|$ is the number
of subscribers who have rated movie $i$), $\rho_i$  and $d_i$ are the
rating and date, respectively, of movie $i$ in the auxiliary information,
and $\rho'_i$  and $d'_i$ are the rating and date in the candidate
record $r'$.  As explained in section~\ref{meta-alg}, this scoring
function was chosen to favor statistically unlikely matches and thus
minimize accidental false positives.  The parameters $\rho_0$ and $d_0$
are 1.5 and 30, days respectively.  These were chosen heuristically,
as they gave the best results in our experiments.  The same parameters
were used throughout, regardless of the amount of noise in $\Aux$.
The eccentricity parameter was set to $\phi=1.5$, \ie, the algorithm
declares there is no match if and only if the difference between the
highest and the second highest scores is no more than 1.5 times the
standard deviation. (A constant value of the eccentricity does not always
give the equal error rate, but it is a close enough approximation.)

\vspace{1ex}
\noindent
\textbf{Didn't Netflix publish only a sample of the data?}
Because Netflix published only $\frac{1}{8}$ of its 2005 database,
we need to be concerned about false positives.  What if the adversary
finds a record matching his $\aux$ in the published sample, but this is
a false match and the ``real'' record has not been released at all?

\begin{figure}
\begin{minipage}[b]{0.47\linewidth} 
\centering
\mbox{\epsfig{file=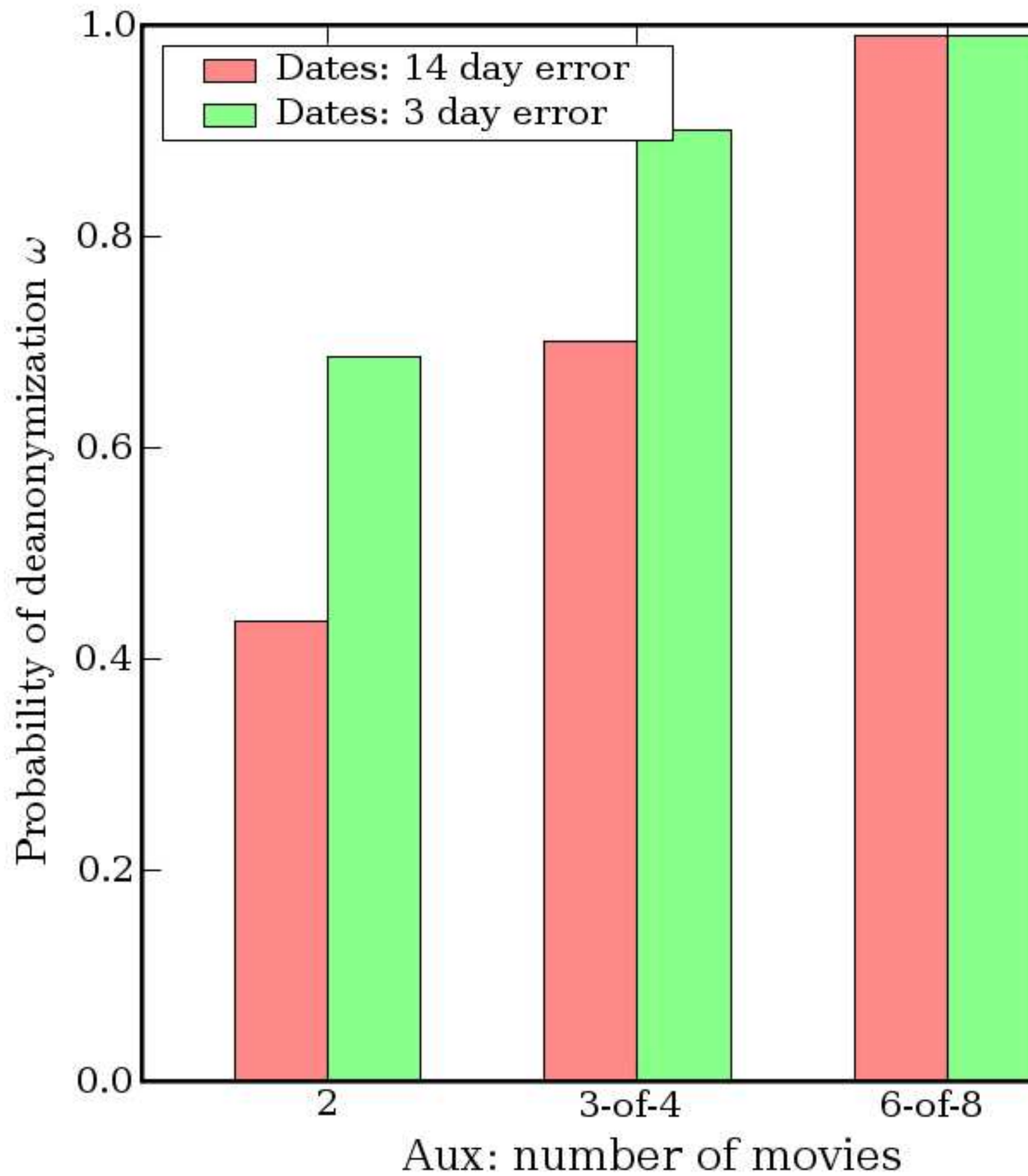,width=3in}}
\caption{De-anonymization: adversary knows exact ratings and approximate
dates.\label{fig:erp}}
\end{minipage}
\hspace{.5cm} 
\begin{minipage}[b]{0.47\linewidth}
\centering
\mbox{\epsfig{file=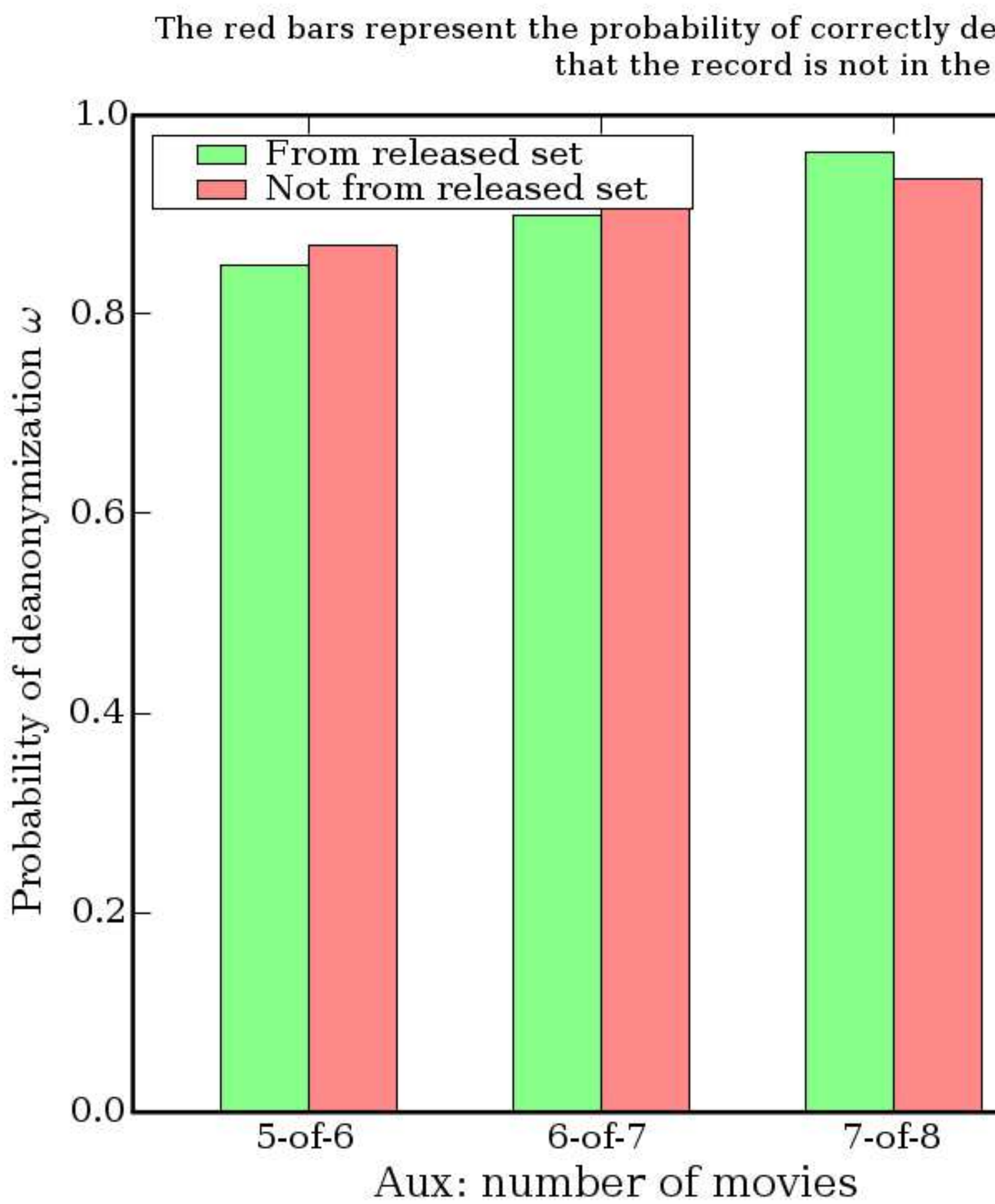,width=3in}}
\caption{Same parameters as Figure~\ref{fig:erp}, but the adversary
is also required to detect when the target record is not in the
sample.\label{fig:eqp}}
\end{minipage}
\end{figure}

Our algorithm is specifically designed to detect when the record
corresponding to $\aux$ is \emph{not} in the sample.  To verify this, we
ran the following experiment.  First, we gave $\aux$ from a random record
to the algorithm and ran it on the dataset.  Then we \emph{removed} this
record from the dataset and re-ran the algorithm.  In the former case,
we expect the algorithm to find the record; in the latter, to declare
that the record is not in the dataset.  As shown in Figure~\ref{fig:eqp},
the algorithm succeeds with high probability in both cases.

It is possible, although \emph{extremely} unlikely, that the original
Netflix dataset is \emph{not} as sparse as the published sample, \ie, it
contains clusters of records which are close to each other, but only one
representative of each cluster has been released in the Prize dataset.
A dataset with such a structure would be exceptionally unusual and
theoretically problematic (see Theorem~\ref{thmsample}).

Finally, our de-anonymization algorithm is still useful even if the amount
of auxiliary information available to the adversary is less than shown
in Figure~\ref{fig:eqp}.  While the absence of false positives cannot
be guaranteed \emph{a priori} in this case, there is a lot of additional
information in the dataset that can be used to eliminate false positives.
For example, consider the start date and the total number of movies in
a record.  If these are part of the auxiliary information (\eg, the
adversary knows approximately when his target first joined Netflix),
they can be used to eliminate candidate records.

\begin{figure}
\begin{minipage}[b]{0.47\linewidth} 
\centering
\mbox{\epsfig{file=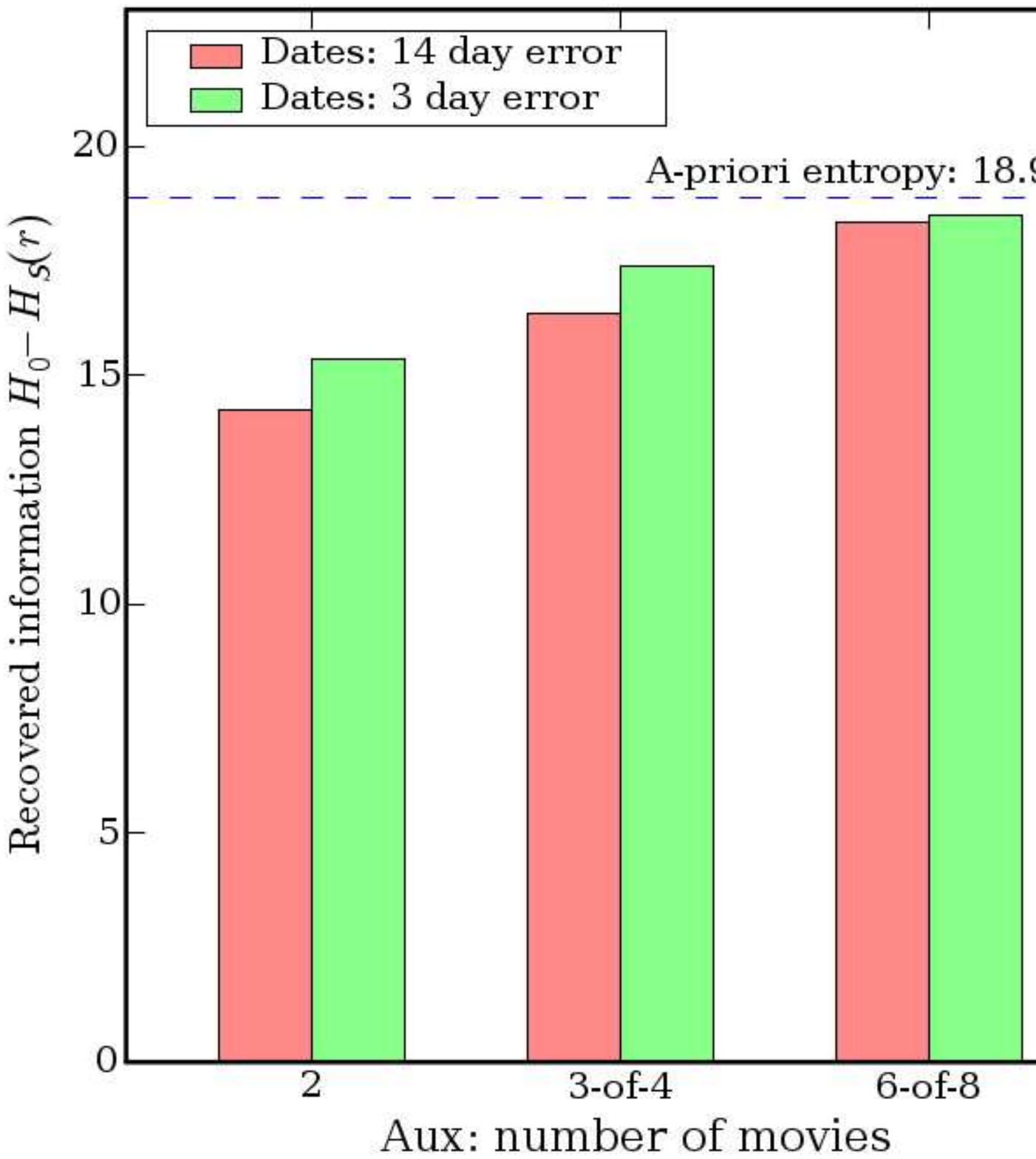,width=3in}}
\caption{Entropic de-anonymization: 
same parameters as in Figure~\ref{fig:erp}.\label{fig:ere}}
\end{minipage}
\hspace{.5cm} 
\begin{minipage}[b]{0.47\linewidth}
\centering
\mbox{\epsfig{file=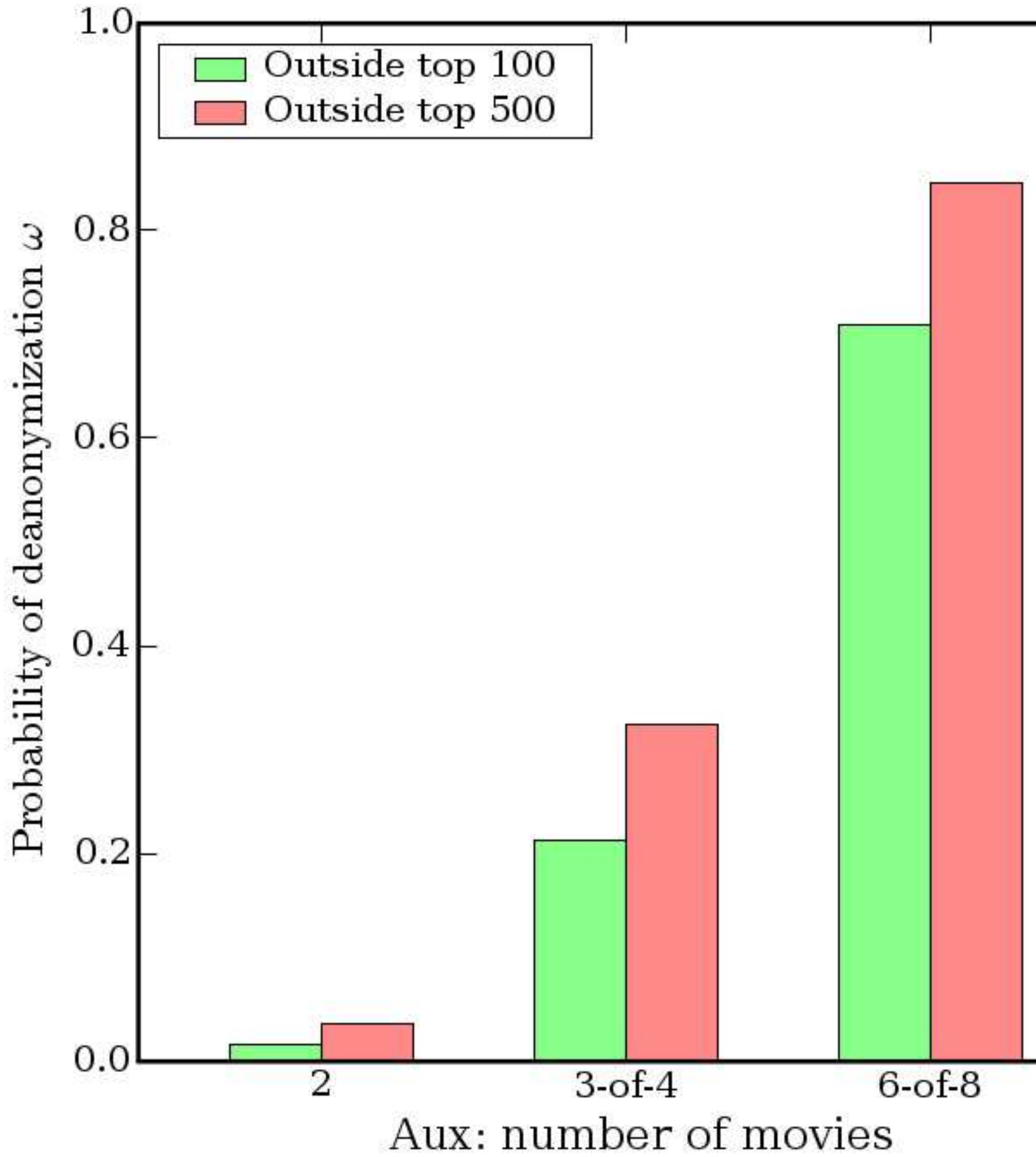,width=3in}}
\caption{Adversary knows exact ratings but does not know dates at
all.\label{fig:ndp}}
\end{minipage}
\end{figure}

\vspace{1ex}
\noindent
\textbf{Results of de-anonymization.}
Very little auxiliary information is needed for de-anonymize an
average subscriber record from the Netflix Prize dataset.  With 8 movie
ratings (of which 2 may be completely wrong) and dates that may have
a 14-day error, 99\% of records be uniquely identified in the dataset.
For 68\%, \emph{two} ratings and dates (with a 3-day error) are sufficient
(Figure~\ref{fig:erp}).  Even for the other 32\%, the number of possible
candidates is brought down dramatically.  In terms of entropy, the
additional information required for complete de-anonymization is only
around 3 bits in the latter case (with no auxiliary information, this
number is 19 bits).  When the adversary knows 6 movies correctly and 2
incorrectly, the extra information he needs for complete de-anonymization
is a fraction of a bit (Figure~\ref{fig:ere}).

Even without any dates, a substantial privacy breach occurs,
especially when the auxiliary information consists of movies that are
not blockbusters (Figure \ref{fig:ndp}).\footnote{We measure the rank of
a movie by the number of subscribers who have rated it.}  Two movies are
no longer sufficient, but 84\% of subscribers can be uniquely identified
if the adversary knows 6 out of 8 moves outside the top 500 (as shown
in appendix~\ref{marginals}, this is not a significant limitation).

\begin{figure}
\begin{minipage}[b]{0.47\linewidth} 
\centering
\mbox{\epsfig{file=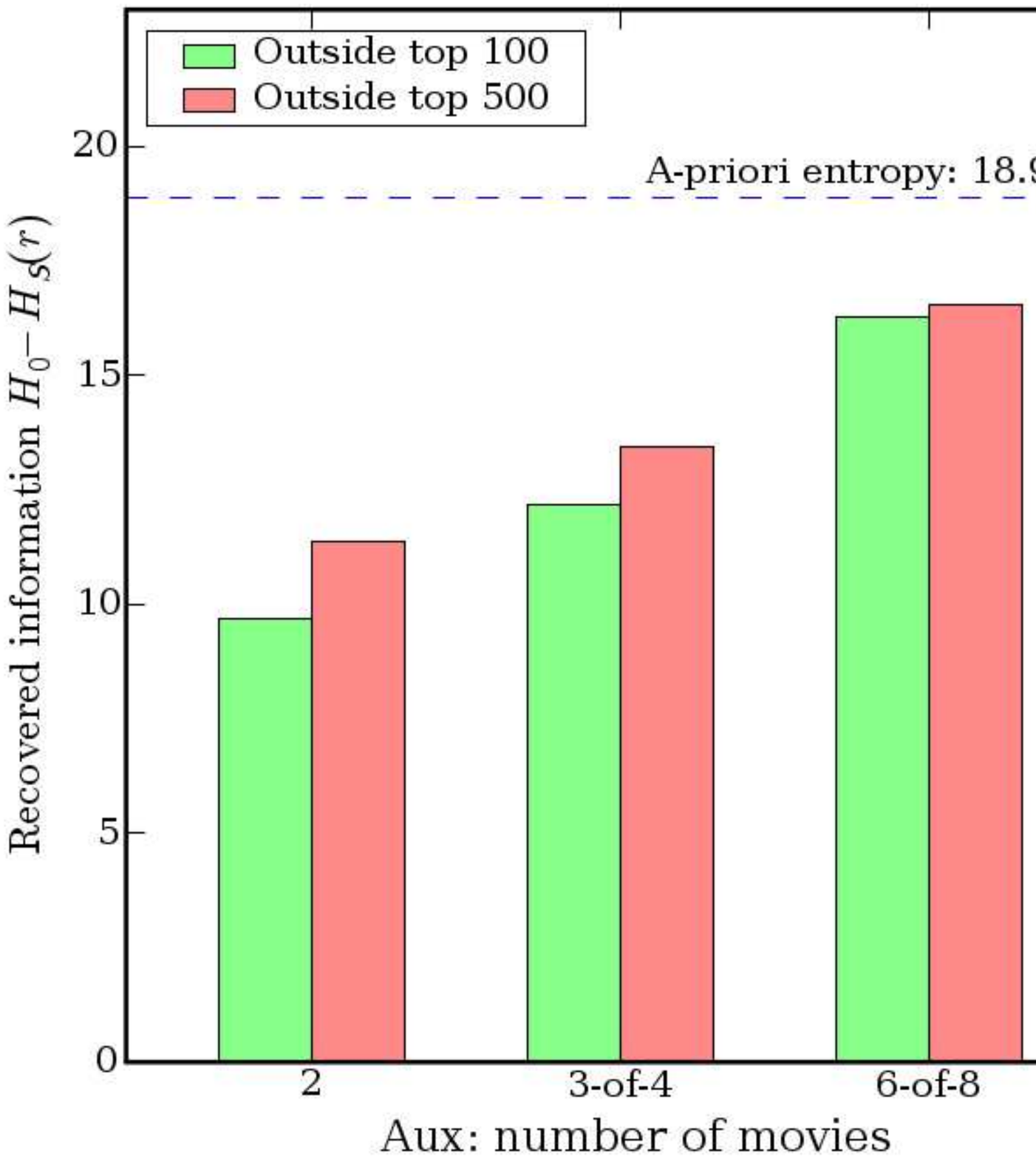,width=3in}}
\caption{Entropic de-anonymization: 
same parameters as in Figure~\ref{fig:ere}.\label{fig:nde}}
\end{minipage}
\hspace{.5cm} 
\begin{minipage}[b]{0.47\linewidth}
\centering
\mbox{\epsfig{file=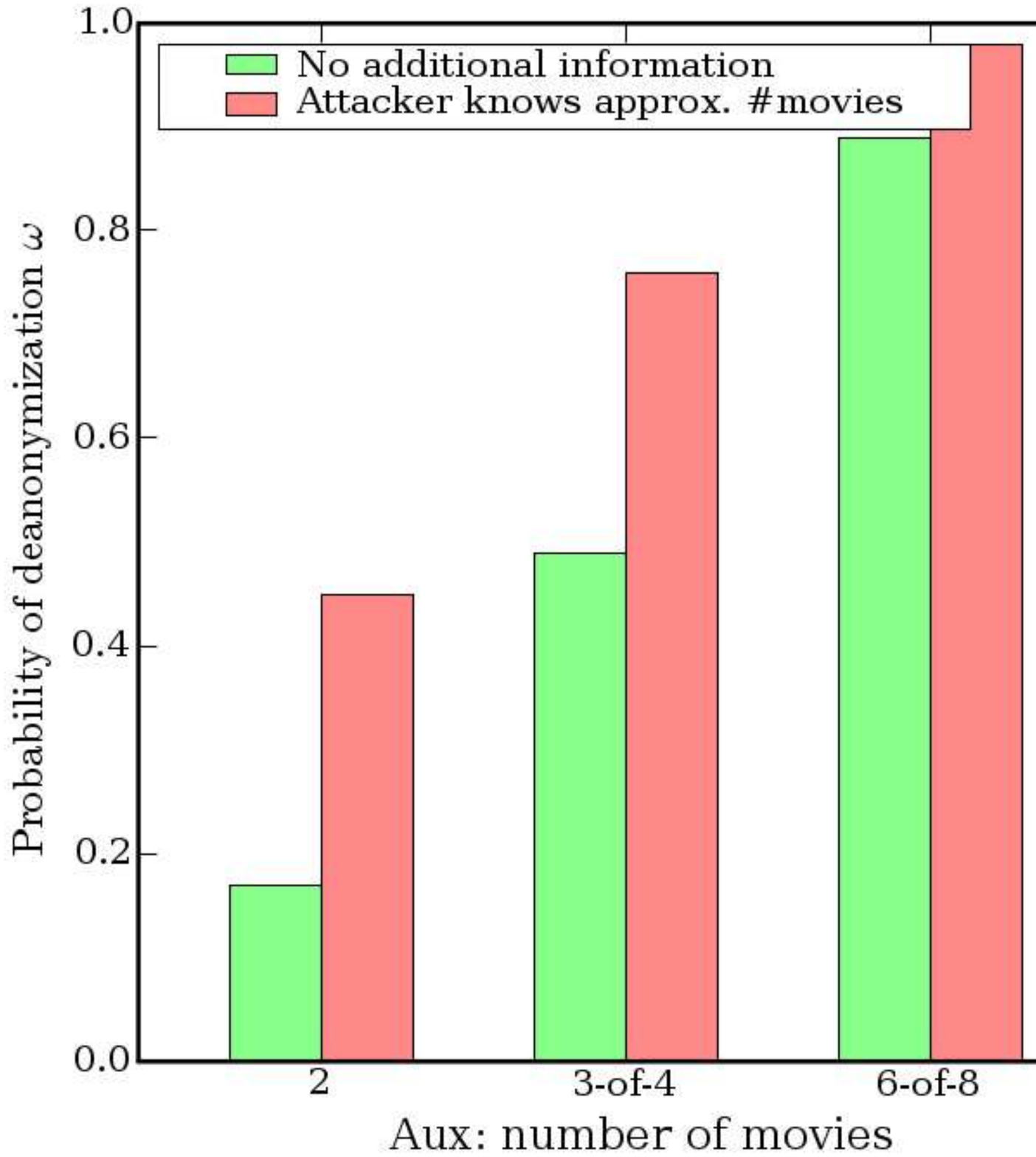,width=3in}}
\caption{Effect of knowing approximate number of movies rated by victim
($\pm 50\%$.)  Adversary knows approximate ratings($\pm 1$) and dates
(14-day error).\label{fig:clp}}
\end{minipage}
\end{figure}

Figure~\ref{fig:nde} shows that even when the adversary's probability
to correctly determine the attributes of the target record is low,
he gains a tremendous amount of information about the target record.
Even in the most pessimistic scenario, the additional information he
would need to complete the de-anonymization has been reduced to less
than half of its original value.

Figure \ref{fig:clp} shows why even partial de-anonymization can be very
dangerous.  There are many things the adversary might know about his
target that are not captured by our formal model, such as the approximate
number of movies rated, the date when they joined Netflix and so on.
Once a candidate set of records is available, further automated analysis
or human inspection might be sufficient to complete the de-anonymization.
Figure~\ref{fig:clp} shows that in some cases, knowing the number of
movies the target has rated (even with a 50\% error!) can more than
double the probability of complete de-anonymization.

\vspace{1ex}
\noindent
\textbf{Obtaining the auxiliary information and IMDb cross-correlation.}
\label{imdb}
Given how little auxiliary information is needed to de-anonymize the
average subscriber record from the Netflix Prize dataset, a determined
adversary should not find it difficult to obtain such information,
especially since it need not be precise.  A water-cooler conversation with
an office colleague about her cinematographic likes and dislikes may yield
enough information, especially if at least a few of the movies mentioned
are outside the top 100 most rated Netflix movies.  This information
can also be gleaned from personal blogs, Google searches, and so on.

One possible source of users' movie ratings is the Internet Movie Database
(IMDb)~\cite{imdb}.  We expect that for Netflix subscribers who use IMDb,
there is a strong correlation between their private Netflix ratings and
their public IMDb ratings.  Note that our attack does \emph{not} require
that all movies rated by the subscriber in the Netflix system be also
rated in IMDb, or vice versa.  In many cases, even a handful of movies
that are rated by the subscriber in both services would be sufficient
to identify his or her record in the Netflix Prize dataset, if present
among the released records, with enough statistical confidence to rule
out the possibility of a false match except for a negligible probability.


Due to the restrictions on crawling IMDb imposed by IMDb's terms
of service (of course, a real adversary may not comply with these
restrictions), we worked with a very small sample of a few dozen IMDb
users.  Results presented in this section should thus be viewed as a
proof of concept.  They do not imply anything about the percentage of
IMDb users who can be identified in the Netflix Prize dataset.

The auxiliary information obtained from IMDb is quite noisy.  First,
a significant fraction of the movies rated on IMDb are not in Netflix,
and vice versa, \eg, movies that have not been released in the US.
Second, some of the ratings on IMDb are missing (\ie, the user entered
only a comment, not a numerical rating).  Such data are still useful for
de-anonymization because an average user has rated only a tiny fraction
of all movies, so the mere fact that a person has watched a given
movie tremendously reduces the number of anonymous Netflix records that
could possibly belong to that user.  Finally, IMDb users among Netflix
subscribers fall into a continuum of categories with respect to rating
dates, separated by two extremes: some meticulously rate movies on both
IMDb and Netflix at the same time, and others rate them whenever they
have free time (which means the dates may not be correlated at all).
Somewhat offsetting these disadvantages is the fact that we can use all
of the user's ratings publicly available on IMDb.

Because we have no ``oracle'' to tell us whether the record our algorithm
has found in the Netflix Prize dataset based on the ratings of some IMDb
user indeed belongs to that user, we need to guarantee a very low false
positive rate.  Given our small sample of IMDb users (recall that it
was deliberately kept small to comply with IMDb's terms of service),
our algorithm identified the records of two users the Netflix Prize
dataset with eccentricities of around 28 and 15, respectively.  This is
an exceptionally strong match.  The records in questions are \textbf{28
standard deviations} (respectively, 15 standard deviations) away from the
second-best candidate.  Interestingly, the first user was de-anonymized
mainly from the ratings and the second mainly from the dates.  Also,
for nearly all the other IMDb users we tested, the eccentricity was no
more than 2.


Let us summarize what our algorithm achieves.  Given a user's
\emph{public} IMDb ratings, which the user posted voluntarily to
selectively reveal \emph{some} of his (or her; but we'll use the male
pronoun without loss of generality) movie likes and dislikes, we discover
\emph{all} the ratings that he entered \emph{privately} into the Netflix
system, presumably expecting that they will remain private.  A natural
question to ask is why would someone who rates movies on IMDb---often
under his or her real name---care about privacy of his movie ratings?
Consider the information that we have been able to deduce by locating
one of these users' entire movie viewing history in the Netflix dataset
and that \emph{cannot} be deduced from his public IMDb ratings.

First, we can immediately find his political orientation based on his
strong opinions about ``Power and Terror: Noam Chomsky in Our Times''
and ``Fahrenheit 9/11.'' Strong guesses about his religious views can be
made based on his ratings on ``Jesus of Nazareth'' and ``The Gospel of
John''. He did not like ``Super Size Me'' at all; perhaps this implies
something about his physical size?  Both items that we found with
predominantly gay themes, ``Bent'' and ``Queer as folk'' were rated
one star out of five.  He is a cultish follower of ``Mystery Science
Theater 3000''. This is far from all we found about this one person,
but having made our point, we will spare the reader further lurid details.

\section{Conclusions}

We have presented a de-anonymization methodology for multi-dimensional
micro-data, and demonstrated its practical applicability by showing how to
de-anonymize movie viewing records released in the Netflix Prize dataset.
Our de-anonymization algorithm works under very general assumptions
about the distribution from which the data are drawn, and is robust
to perturbation and sanitization.  Therefore, we expect that it can
be successfully used against any large dataset containing anonymous
multi-dimensional records such as individual transactions, preferences,
and so on.

An interesting topic for future research is extracting
social relationships, networks and clusters from the anonymous
records.  This knowledge can be a source of information for further
de-anonymization~\cite{hayes}.  In the case of the Netflix Prize dataset,
de-anonymization of individual records may also have interesting
implications for winning the Netflix Prize.  We discuss this briefly
in appendix~\ref{winprize}.

\bibliographystyle{plain}
\bibliography{netflix}

\begin{thebibliography}{10}

\bibitem{AW89}
N.~Adam and J.~Worthmann.
\newblock Security-control methods for statistical databases: {A} comparative
  study.
\newblock {\em {ACM} Computing Surveys}, 21(4):515--556, 1989.

\bibitem{curse}
C.~Aggarwal.
\newblock On k-anonymity and the curse of dimensionality.
\newblock In {\em Proc. 31st International Conference on Very Large Data Bases
  ({VLDB})}, pages 901--909. {ACM}, 2005.

\bibitem{AS00}
R.~Agrawal and R.~Srikant.
\newblock Privacy-preserving data mining.
\newblock In {\em Proc.\ 2000 {ACM} {SIGMOD} International Conference on
  Management of Data}, pages 439--450. {ACM}, 2000.

\bibitem{longtail}
C.~Anderson.
\newblock {\em The Long Tail: Why the Future of Business Is Selling Less of
  More}.
\newblock Hyperion, 2006.

\bibitem{sulq}
A.~Blum, C.~Dwork, F.~McSherry, and K.~Nissim.
\newblock Practical privacy: {The} {SuLQ} framework.
\newblock In {\em Proc.\ 24th {ACM} {SIGACT}-{SIGMOD}-{SIGART} Symposium on
  Principles of Database Systems ({PODS})}, pages 128--138. {ACM}, 2005.

\bibitem{bryn}
E.~Brynjolfsson, Y.~Hu, and M.~Smith.
\newblock Consumer surplus in the digital economy.
\newblock {\em Management Science}, 49(11), 2003.

\bibitem{CDM05}
S.~Chawla, C.~Dwork, F.~{McSherry}, A.~Smith, and H.~Wee.
\newblock Towards privacy in public databases.
\newblock In {\em Proc.\ 2nd Theory of Cryptography Conference ({TCC})}, volume
  3378 of {\em LNCS}, pages 363--385. Springer, 2005.

\bibitem{ciriani}
V.~Ciriani, S.~De~Capitani di~Vimercati, S.~Foresti, and P.~Samarati.
\newblock k-anonymity.
\newblock {\em Secure Data Management in Decentralized Systems}, 2007.

\bibitem{DSCP02}
C.~D{\'{\i}}az, S.~Seys, J.~Claessens, and B.~Preneel.
\newblock Towards measuring anonymity.
\newblock In {\em Proc.\ 2nd International Workshop on Privacy-Enhancing
  Technologies}, volume 2482 of {\em LNCS}, pages 54--68, 2002.

\bibitem{lens}
D.~Frankowski, D.~Cosley, S.~Sen, L.~Terveen, and J.~Riedl.
\newblock You are what you say: privacy risks of public mentions.
\newblock In {\em Proc.\ 29th Annual {ACM} {SIGIR} Conference on Research and
  Development in Information Retrieval}, pages 565--572. {ACM}, 2006.

\bibitem{nyt}
K.~Hafner.
\newblock And if you liked the movie, a {Netflix} contest may reward you
  handsomely.
\newblock New York Times, Oct 2 2006.

\bibitem{aol}
S.~Hansell.
\newblock {AOL} removes search data on vast group of web users.
\newblock New York Times, Aug 8 2006.

\bibitem{hayes}
B.~Hayes.
\newblock Connecting the dots: Can the tools of graph theory and social-network
  studies unravel the next big plot?
\newblock {\em American Scientist}, September--October 2006.
\newblock
  \url{http://www.americanscientist.org/template/AssetDetail/assetid/53062}.

\bibitem{imdb}
{IMDb}.
\newblock {The Internet Movie Database}.
\newblock \url{http://www.imdb.com/}, 2007.

\bibitem{jensen}
J.~L. W.~V. Jensen.
\newblock Sur les fonctions convexes et les inégalités entre les valeurs
  moyennes.
\newblock {\em Acta Mathematica}, 30(1):175--193, 1906.

\bibitem{LAH06}
J.~Leskovec, L.~Adamic, and B.~Huberman.
\newblock The dynamics of viral marketing.
\newblock In {\em Proc.\ 7th {ACM} Conference on Electronic Commerce}, pages
  228--237. ACM, 2006.

\bibitem{ldiversity}
A.~Machanavajjhala, J.~Gehrke, D.~Kifer, and M.~Venkitasubramaniam.
\newblock l-diversity: Privacy beyond k-anonymity.
\newblock In {\em Proc.\ 22nd International Conference on Data Engineering
  ({ICDE})}. {IEEE Computer Society}, 2006.

\bibitem{icde07}
A.~Machanavajjhala, D.~Martin, D.~Kifer, J.~Gehrke, and J.~Halpern.
\newblock Worst case background knowledge.
\newblock In {\em Proc.\ 23rd International Conference on Data Engineering
  ({ICDE})}. {IEEE Computer Society}, 2007.

\bibitem{malin}
B.~Malin and L.~Sweeney.
\newblock How (not) to protect genomic data privacy in a distributed network:
  using trail re-identification to evaluate and design anonymity protection
  systems.
\newblock {\em J. of Biomedical Informatics}, 37(3):179--192, 2004.

\bibitem{faq}
Netflix.
\newblock {Netflix Prize: FAQ}.
\newblock \url{http://www.netflixprize.com/faq}, Downloaded on Oct 17 2006.

\bibitem{SD02}
A.~Serjantov and G.~Danezis.
\newblock Towards an information theoretic metric for anonymity.
\newblock In {\em Proc.\ 2nd International Workshop on Privacy Enhancing
  Technologies ({PET})}, volume 2482 of {\em LNCS}, pages 41--53. Springer,
  2003.

\bibitem{sweeney}
L.~Sweeney.
\newblock Weaving technology and policy together to maintain confidentiality.
\newblock {\em J. of Law, Medicine and Ethics}, 25(2--3):98--110, 1997.

\bibitem{sweeney-supp}
L.~Sweeney.
\newblock Achieving k-anonymity privacy protection using generalization and
  suppression.
\newblock {\em International J. of Uncertainty, Fuzziness and Knowledge-based
  Systems}, 10(5):571--588, 2002.

\bibitem{ksweeney}
L.~Sweeney.
\newblock k-anonymity: A model for protecting privacy.
\newblock {\em International J. of Uncertainty, Fuzziness and Knowledge-based
  Systems}, 10(5):557--570, 2002.

\bibitem{collab}
J.~Thornton.
\newblock Collaborative filtering research papers.
\newblock \url{http://jamesthornton.com/cf/}, 2006.

\bibitem{TYW84}
J.~Traub, Y.~Yemini, and H.~Wozniakowski.
\newblock The statistical security of a statistical database.
\newblock {\em {ACM} Transactions on Database Systems}, 9(4):672--679, 1984.

\bibitem{leizhang}
L.~Zhang, S.~Jajodia, and A.~Brodsky.
\newblock Information disclosure under realistic assumptions: Privacy versus
  optimality.
\newblock In {\em Proc.\ 14th {ACM} Conference on Computer and Communications
  Security ({CCS})}. {ACM}, 2007.

\end{thebibliography}

\begin{appendix}
\section{Glossary of terms}

\begin{tabular}{|l|l|}
\hline
Symbol&Meaning \\
$D$&Database\\
${\hat D}$&Released sample\\
$N$&Number of rows\\
$M$&Number of columns\\
$m$&Size of $\aux$\\
$X$&Domain of attributes\\
$\perp$&Null attribute\\
$\supp(.)$&Set of non-null attributes in a row/column\\
$\Sim$&Similarity measure\\
$\Aux$&Auxiliary information sampler\\
$\aux$&Auxiliary information\\
$\score$&Scoring function\\
$\epsilon$&Sparsity threshold\\
$\delta$&Sparsity probability\\
$\theta$&Closeness of de-anonymized record\\
$\omega$&Probability of success of de-anonymization\\
$r, r'$&Record\\
$\Pi$&P.d.f over records\\
$H_S$&Shannon entropy\\
$H$&De-anonymization entropy\\
$\phi$&Eccentricity\\
\hline
\end{tabular}

\section{Implications for the Netflix Prize}
\label{winprize}

\begin{figure}[h]
\centering
\mbox{\epsfig{file=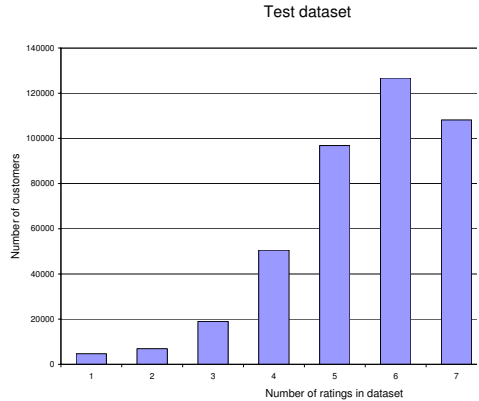,width=0.45\textwidth}}
\caption{Test dataset for the Netflix Prize.
\label{fig:testdata}}
\end{figure}

De-anonymization of Netflix subscribers may enable one to learn the
true ratings for some entries in the Netflix Prize \emph{test} dataset
(these ratings have been kept secret by Netflix).  The test dataset has
been chosen in such a way that the contribution of any given subscriber
is no more than 9 entries (see Figure~\ref{fig:testdata}).  Therefore,
it is not possible to find a small fraction of subscribers whose ratings
will reveal a large fraction of the test dataset.

Access to true ratings on the test dataset does not translate to an
immediate strategy for claiming the Netflix Prize.  The rules require
that the algorithm be submitted for perusal.  In spite of this, having the
test data (or the data closely correlated with the test data) enables the
contestant to train on the test data in order to ``overfit'' the model.
This is why Netflix kept the ratings on the test data secret.

How many Netflix subscribers would need to be de-anonymized before there
is a significant impact on the performance of a recommendation algorithm
that uses this information?  The root mean squared errors (RMSE) of
the current top performers (as of November 9, 2007) are about $0.87$.  If a
subscriber's ``true'' ratings are available, the error for that subscriber
drops to zero.  Thus, if the learner has access to $\frac{1}{0.87} =
1.14\%$ of de-anonymized records, then the RMSE score improves by 1\%
(assuming that the contribution of each subscriber is the same and RMSE
behaves roughly linearly).  This is roughly equal to the difference the
current 1st and 20th contestants on the Netflix Prize leader board.

How easy is it to de-anonymize 1\% of the subscribers? The potential
sources of large-scale true rating data are the publicly available
ratings on the site itself, public ratings on other movie websites such
as IMDb, and the subscribers themselves.  Netflix appears to have taken
the elementary precaution of removing from the dataset the ratings of
the subscribers that are publicly available on the Netflix website.
While our experiments in section~\ref{imdb} show that successful
cross-correlation of IMDb and Netflix records is possible, there are
some hurdles to overcome: it is not clear what fraction of users with a
significant body of movie ratings on IMDb are also Netflix subscribers,
nor is it known how ratings and dates on IMDb correlate with those on
Netflix for the average user (although we expect a strong correlation).

Collecting data from the subscribers themselves appears to be the most
promising direction.  Many Netflix subscribers do not regard their
ratings as private data and are eager to share them, to the extent
that there even exists a browser plugin that automates this process,
although we have not found any public rating lists generated this way.
If the here-are-all-my-Netflix-ratings ``meme'' propagates through the
``blogosphere,'' it could easily result in a publicly available dataset of
sufficiently large size.  It is also easy for a malicious person to bribe
subscribers (say, ``upload your Netflix ratings to gain access to the
protected areas of this site'').  Also, many subscribers have ``friends''
on Netflix, and subscribers' ratings are accessible to their friends.

We emphasize that even though many Netflix subscribers do not regard
their movie viewing histories as sensitive, this does \emph{not} mean
that privacy of Netflix records is moot.  In section~\ref{imdb}, we
extracted from the Netflix Prize dataset non-public information about
some subscribers that should be considered sensitive by any reasonable
definition.

\section{On perturbation of the Netflix Prize dataset}
\label{liars}

\begin{figure}[h]
\begin{minipage}[b]{0.47\linewidth}
\centering
\mbox{\epsfig{file=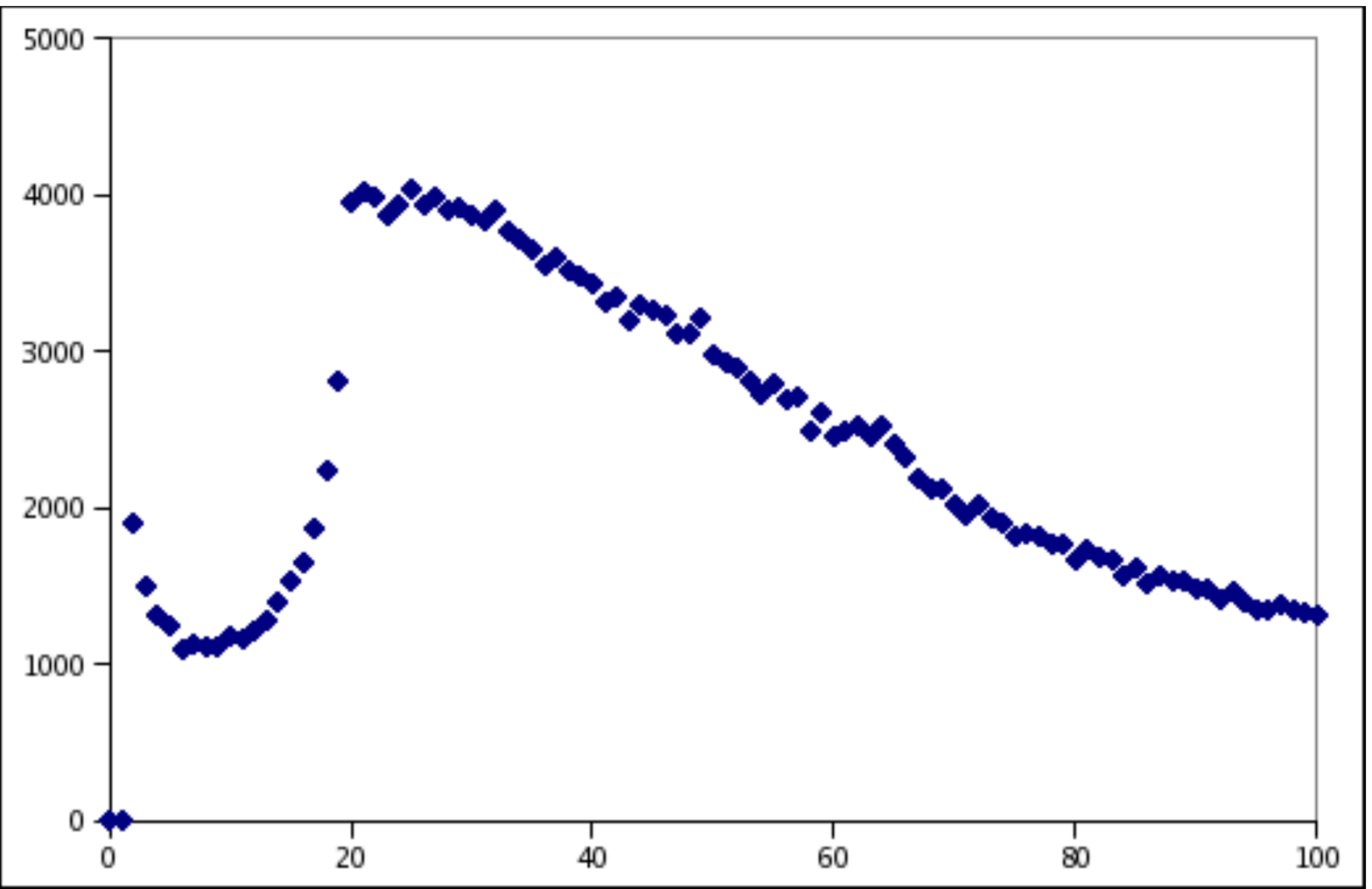,width=0.8\textwidth}}
\caption{For each $k \leq 100$, the number of subscribers with $k$ ratings in
the released dataset.
\label{fig:stats100}}

\end{minipage}
\begin{minipage}[b]{0.47\linewidth}
\centering
\mbox{\epsfig{file=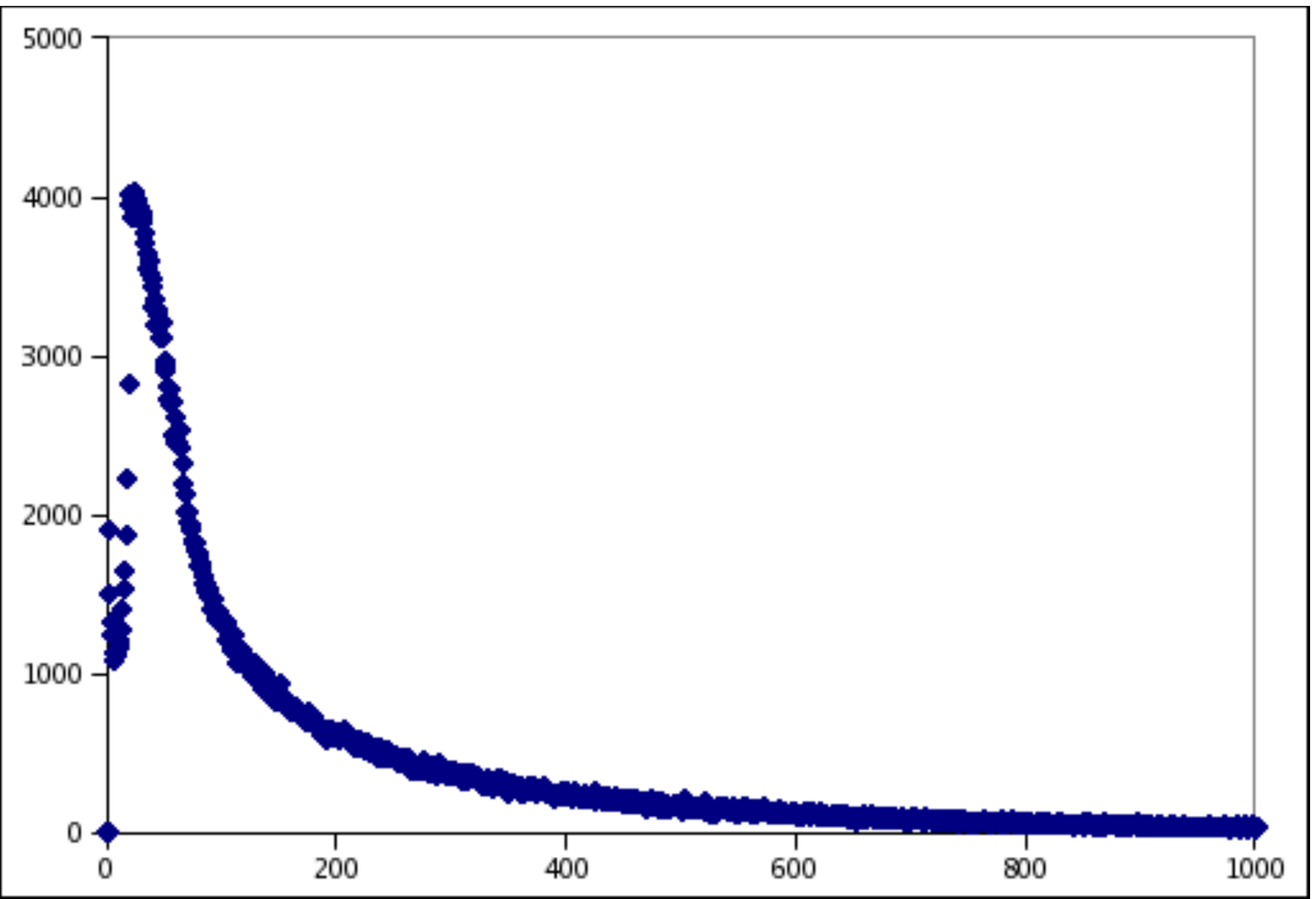,width=0.8\textwidth}}
\caption{For each $k \leq 1000$, the number of subscribers with $k$ ratings in
the released dataset.
\label{fig:stats1000}}
\end{minipage}
\end{figure}

Figs.~\ref{fig:stats100} and~\ref{fig:stats1000} plot the number of
ratings $X$ against the number of subscribers in the released dataset who
have at least $X$ ratings.  The tail is surprisingly thick: thousands of
subscribers have rated more than a thousand movies.  Netflix claims that
the subscribers in the released dataset have been ``randomly chosen.''
Whatever the selection algorithm was, it was not uniformly random.
Common sense suggests that with uniform subscriber selection, the curve
would be monotonically decreasing (as most people rate very few movies
or none at all), and that there would be no sharp discontinuities.

It is not clear how the data was sampled.  Our conjecture is that some
fraction of the subscribers with more than 20 ratings were sampled,
and the points on the graph to the left of $X=20$ are the result of some
movies being deleted after the subscribers were sampled.

We requested the rating history as presented on the Netflix website from
some of our acquaintances, and based on this data (which is effectively
drawn from Netflix's \emph{original}, non-anonymous dataset, since we
know the names associated with these records), located two of them
in the Netflix Prize dataset.  Netflix's claim that the data were
perturbed~\cite{faq} does not appear to be borne out.  One of the
subscribers had 1 of 306 ratings altered, and the other had 5 of 229
altered. (These are upper bounds, because they include the possibility
that the subscribers changed the ratings after the 2005 snapshot
that was released was taken.)  In any case, the level of noise is far
too small to affect our de-anonymization algorithms, which have been
specifically designed to withstand this kind of imprecision.  We have no
way of determining how many dates were altered and how many ratings were
deleted, but we conjecture that very little perturbation has been applied.

It is important that the Netflix Prize dataset has been released to
support development of better recommendation algorithms.  A significant
perturbation of individual attributes would have affected cross-attribute
correlations and significantly decreased the dataset's utility for
creating new recommendation algorithms, defeating the entire purpose of
the Netflix Prize competition.

Finally, we observe that the Netflix Prize dataset clearly has \emph{not}
been $k$-anonymized for any value of $k>1$.

\section{Marginals}
\label{marginals}

\begin{figure}
\begin{minipage}[b]{0.47\linewidth}
\centering
\mbox{\epsfig{file=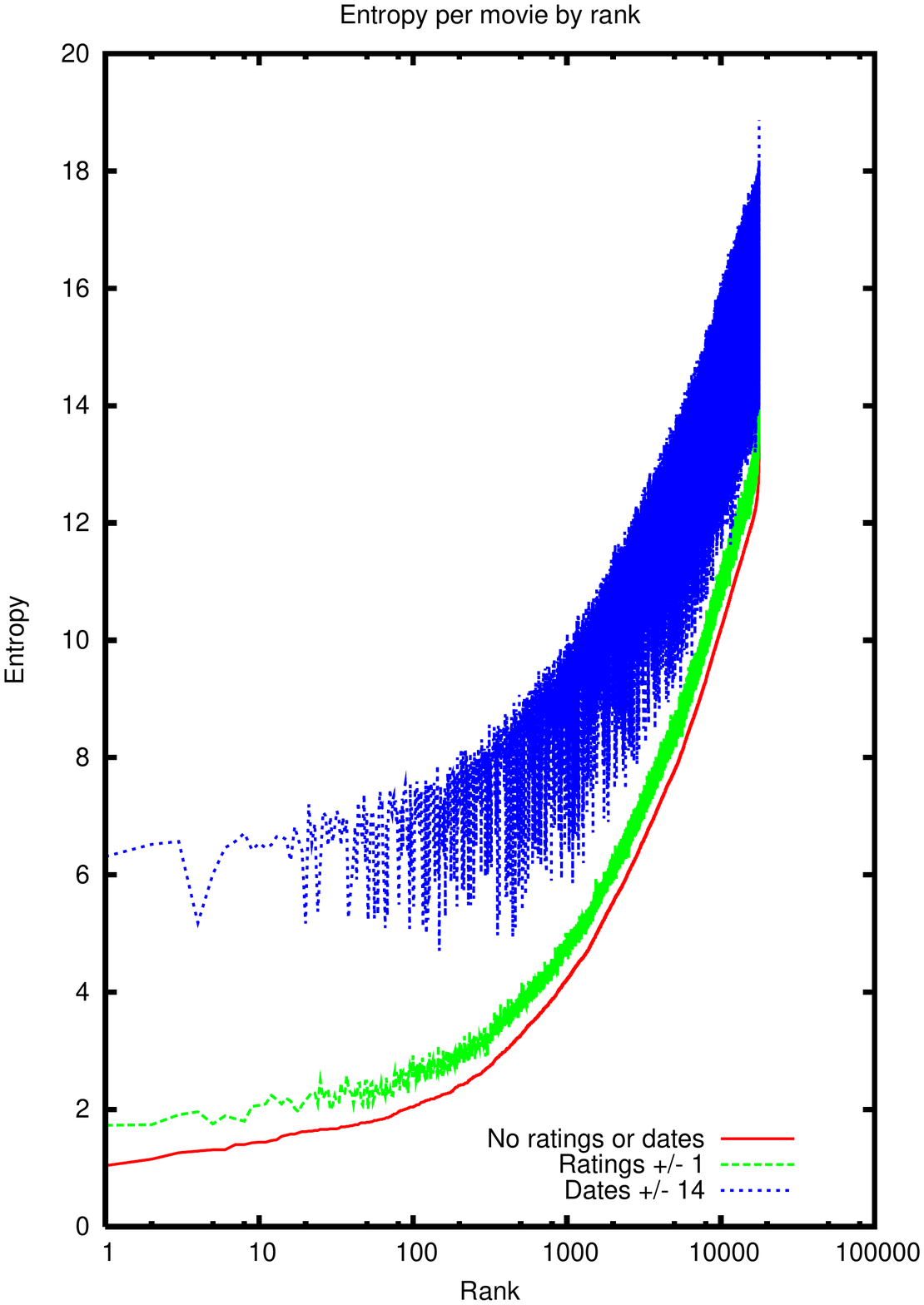,width=0.8\textwidth}}
\caption{Entropy of movie by rank\label{entropyrank}}
\end{minipage}
\begin{minipage}[b]{0.47\linewidth}
\centering
\mbox{\epsfig{file=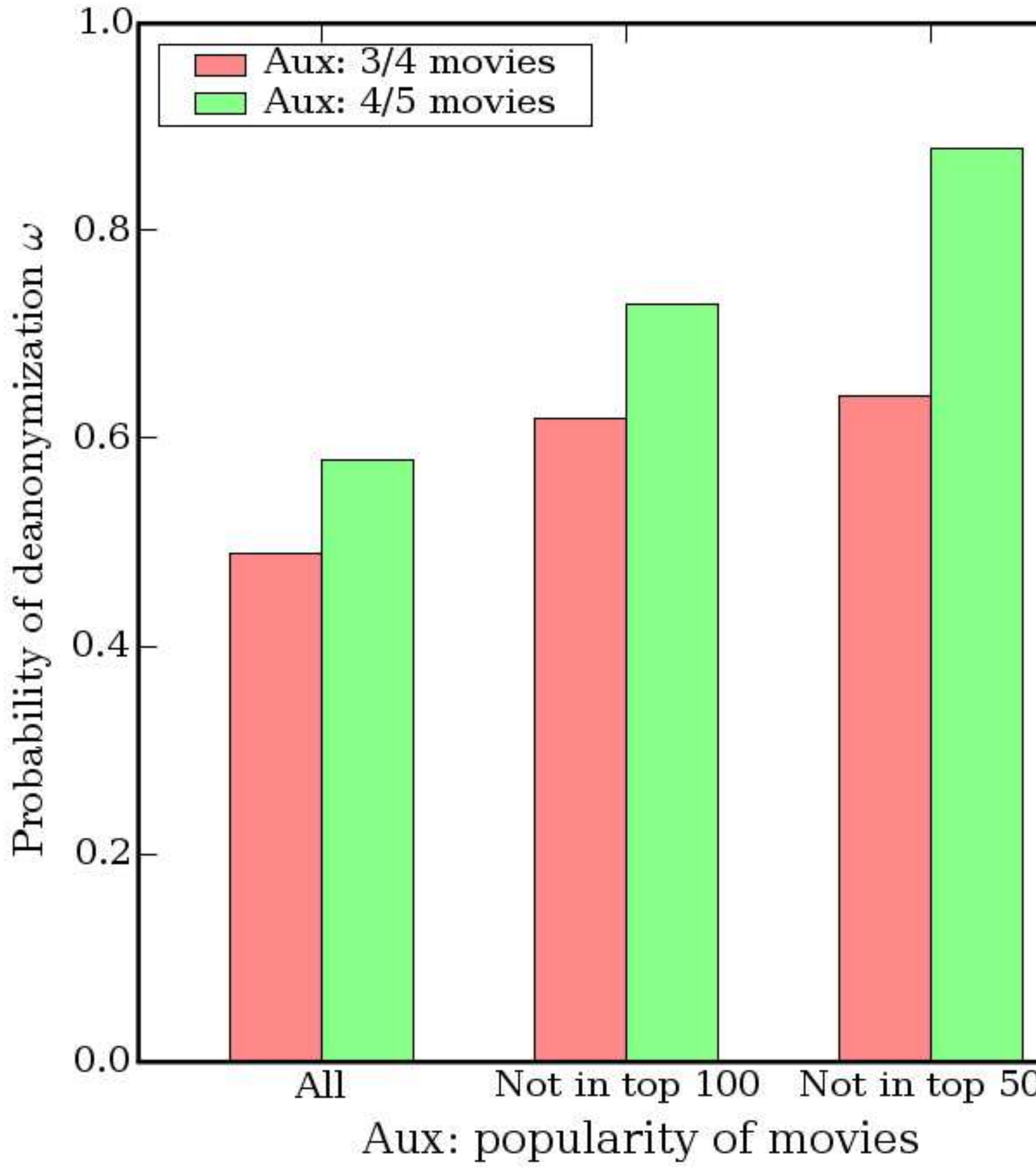,width=3in}}
\caption{Effect of knowing less popular movies rated by victim. Adversary knows approximate ratings ($\pm 1$) and dates (14-day error).} \label{fig:mrp}  
\end{minipage}

\end{figure}

In Figure~\ref{entropyrank}, we demonstrate how much information
the adversary gains about his target from the knowledge of one of
the movies watched by the target, as a function of the rank of the
movie. This helps visualize how the adversary's success varies depending
on whether the movies are randomly picked, or if they are constrained
to be outside the top 100 or top 500.  Of course, since there are
correlations between the lists of subscribers who watched different
movies, we cannot simply multiply the information gain per movie by the
number of movies. Therefore, this graph cannot be used to infer how
many attributes must be part of the auxiliary information before the
adversary can successfuly de-anonymize.

Finally, we show the relationship between subscribers and the ranks of
the movies they rated.

\vspace{1ex}
\begin{tabular}{|r|r|r|r|}
\hline
&
\multicolumn{3}{c|}{
Percentage of subscribers who rated \ldots} \\ \cline{2-4}
& At least 1 movie & At least $5$ & At least $10$ \\
\hline
Not in 100 most rated &
100\% & 97\% & 93\% \\ \hline
Not in 500 most rated &
99\% & 90\% & 80\% \\ \hline
Not in 1000 most rated &
97\% & 83\% & 70\% \\ \hline
\end{tabular}
\vspace{1ex}

Figure~\ref{fig:mrp} explores in greater detail the effect of the
popularity of movies known to the adversary.  The effect is not
negligible, but not dramatic either.

\section{Sparsity}
\label{appendix-sparsity}

The chart demonstrates that most Netflix subscribers do not have even
a \emph{single} subscriber with a high similarity score ($> 0.5$), even
if we consider only the respective sets of movies rated without taking
into account numerical ratings or dates on these movies.

\begin{figure}
\centering
\mbox{\epsfig{file=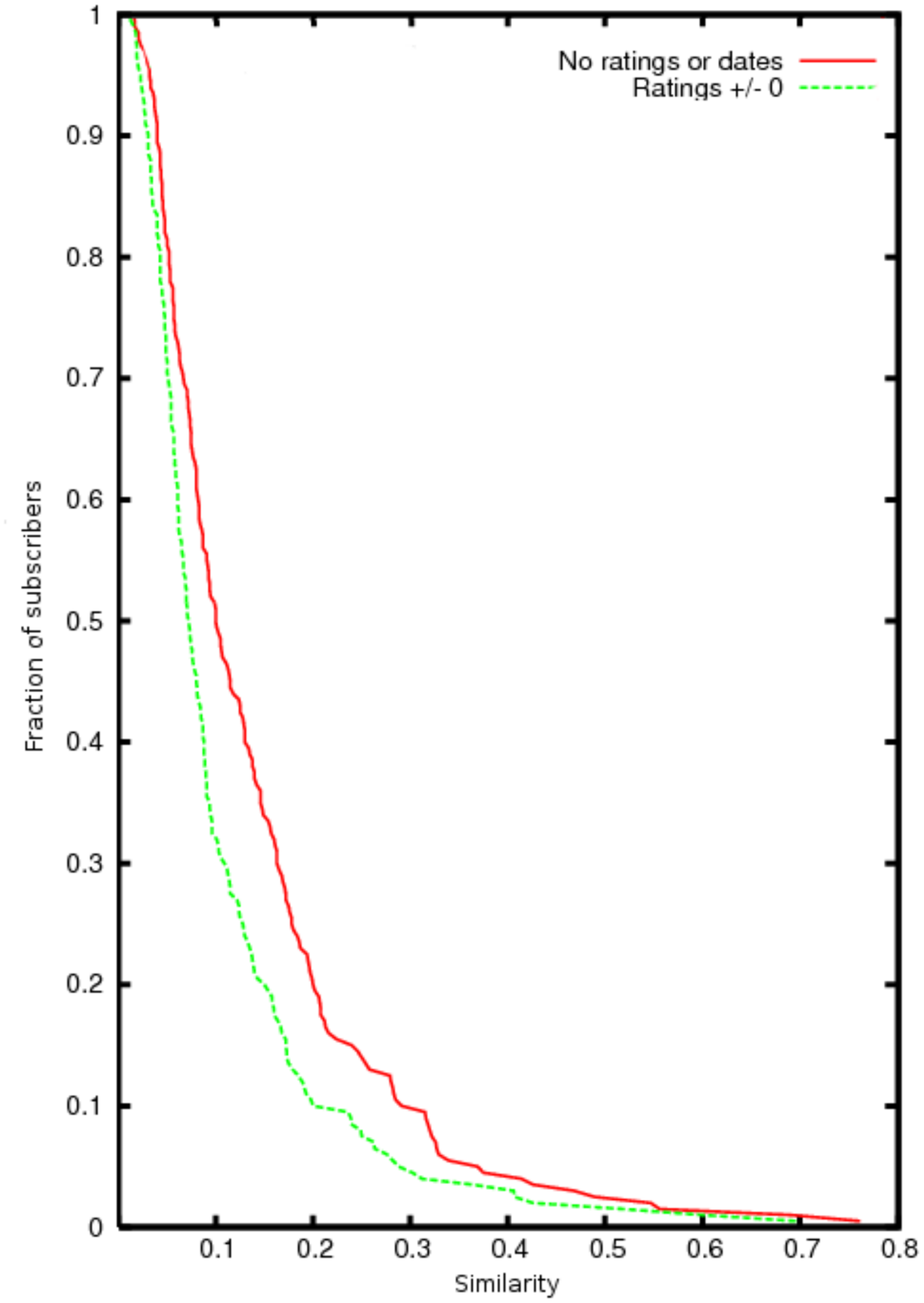,width=0.4\textwidth}}
\caption{X-axis ($x$) is the similarity to nearest neighbor: i.e, the subscriber with the
highest similarity score. Y-axis is the fraction of subscribers whose nearest neighbor
similarity is at least $x$.\label{simchart}}
\centering

\end{figure}

\end{appendix}


\end{document}